\documentclass[preprint,eqsecnum,prb]{revtex4}
\usepackage{epsfig,amsmath,graphicx}

\newcommand{\ee}{\end{equation}}
\newcommand{\be}{\begin{equation}} 
\newcommand{\bea}{\begin{eqnarray}}
\newcommand{\eea}{\end{eqnarray}}
\newcommand{\bml}{\begin{subequations}} 
\newcommand{\eml}{\end{subequations}}

\begin{document}

\title{A multilayer multiconfiguration time-dependent Hartree study of
the nonequilibrium Anderson impurity model at zero temperature}

\author{Haobin Wang}
\affiliation{Department of Chemistry, University of Colorado,
Denver, CO 80217, USA}
\affiliation{Beijing Computational Science Research Center,
No.10 East Xibeiwang Road, Haidian District, Beijing 100193, China}
\author{Michael Thoss}
\affiliation{Institute of Physics, University of Freiburg, Hermann-Herder-Str. 3, D-79104 Freiburg, Germany}

\date{\today}

\begin{abstract}
\baselineskip6mm

Quantum transport is studied for the nonequilibrium Anderson impurity model at zero temperature
employing the multilayer multiconfiguration time-dependent Hartree theory within 
the second quantization representation (ML-MCTDH-SQR) of Fock space.  To adress both linear and nonlinear conductance in the Kondo regime, two new techniques of the ML-MCTDH-SQR simulation methodology are introduced: (i) the use of correlated initial states, which is achieved by imaginary time 
propagation of the overall Hamiltonian at zero voltage and (ii) the adoption of the logarithmic discretization of the electronic continuum.  Employing the improved methodology, the signature of the Kondo effect is analyzed.
\end{abstract}
\maketitle

\section{Introduction}

Nonequilibrium quantum transport in molecular junctions or through quantum dots has received 
much attention recently, both as platform to study fundamental aspects of  a nanoscale many-body quantum system out of equilibrium and due to potential applications in future devices.\cite{ree97:252,joa00:541,Nitzan01,nit03:1384,Cuniberti05,Selzer06,Venkataraman06,Chen07,Galperin08b,Cuevas10}  Experimental observations of interesting transport phenomena, such as, e.g., Coulomb blockade,\cite{par02:722,Heersche06} Kondo 
effect,\cite{lia02:725,Roch09,Parks10,Meded11,Esat16} negative differential resistance,\cite{che99:1550,Gaudioso00,Osorio10} 
switching and hysteresis,\cite{blu05:167,Riel06,Choi06} have inspired extensive theoretical
developments.  One class of theoretical methods is based on concepts such as perturbation theory, factorization ansatzes and/or effective single-particle approaches to tackle the complex many-body problem in an approximate way. 
Examples in the context of transport in molecular junctions include scattering theory,\cite{Bonca95,Ness01,Cizek04,Cizek05,Toroker07,Benesch08,Zimbovskaya09,Seidemann10} 
nonequilibrium Green's function (NEGF) approaches,\cite{Flensberg03,Mitra04,Galperin06,Ryndyk06,Frederiksen07,Tahir08,Haertle08,Bergfield09,Haertle09} 
and master equation methods.\cite{May02,Mitra04,Lehmann04,Pedersen05,Harbola06,Zazunov06,Siddiqui07,Timm08,May08,May08b,Leijnse08,Esposito09,Volkovich11,Haertle11}  The strength of these approaches is that they offer important physical insights into the problems of interest at low to moderate numercial effort. They are also often relatively easy to implement.
Their limitation, on the other hand, is the lack of estimate or control of errors, such that it is often difficult to assess the accuracy of the theoretical predictions.

Thus, the development of another class of methods, which can be systematically converged, i.e. numerically exact methods, is essential to reliably address the difficult issues in nonequilibrium transport, in particular in the strong coupling regime.  These methods include
the numerical path  integral approach,\cite{muh08:176403,wei08:195316,Segal10} real-time 
quantum Monte Carlo simulations,\cite{Werner09,Schiro09,Cohen15} the numerical renormalization 
group approach,\cite{and08:066804} the time-dependent density matrix renormalization 
group approach,\cite{HeidrichMeisner09}, the hierarchical equations of motion
method,\cite{Zheng09,Jiang12,Zheng13,Haertle13c,Haertle14,Schinabeck16} a combination of reduced density matrix
techniques and impurity solvers,\cite{Cohen11,Wilner15} and the multilayer multiconfiguration time-dependent 
Hartree (ML-MCTDH) theory in second quantization representation (SQR) developed by us.\cite{wan09:024114}

In previous work, we have applied the ML-MCTDH-SQR method to several models of nonequilibrium quantum transport, including 
the vibrationally-coupled electron transport model, the Anderson impurity model, and their combination.\cite{Wang11,Wang13,Wilner15,Wang16} 
Our simulations have illustrated important physical effects such as bistability, Coulomb and phonon blockade. Furthermore, employing a transformation of the Hamiltonian to an appropriate 
scattering states representation, we have demonstrated the role of quantum correlation in vibrationally-coupled electron transport,\cite{Wang13b}  which is particularly important in the off-resonant transport regime dominated by cotunneling processes, where approximate approaches such as NEGF or Hartree-Fock predict an incorrectly large current.

In this paper, we employ the ML-MCTDH-SQR method to study the
nonequilibrium Anderson impurity model at zero temperature. In this regime, the Anderson impurity model is known to exhibit Kondo effect.\cite{Hewson93,Wiel00,lia02:725} 
To facilitate the simulation of transport in this regime, 
we introduce two modifications to our existing approach.  The
first is the use of a correlated initial state, obtained by imaginary time propagation
of the full Hamiltonian, to reduce transient peaks in the time-dependent current.  The second
is the use of Wilson's logarithmic discretization of the electronic continuum,\cite{Wilson75} used extensively
in the numerical renormalization group type of methods, to access the electronic states near
the Fermi level of the electrodes.  These two methodology adaptations allow us to address 
new physical regimes that are useful to study the effect of both linear and nonlinear conductance.

The remaining part of the paper is organized as follows. 
Sections~\ref{modeltight}-\ref{mlsqr} outline the physical model, the discretization of the electronic 
continuum, the generation of the a correlated initial state, and the ML-MCTDH-SQR theory, respectively. 
Section~\ref{results} presents results of numerical simulations for a variety of parameter regimes as well as 
an analysis of the transport mechanisms. Section~\ref{conclusions} concludes with a summary.

\section{Model and Observables of Interest}\label{modeltight}

In this paper we consider the single level Anderson impurity model\cite{Anderson61}
that has been investigated in great detail both in equilibrium and 
nonequilibrium.\cite{Hewson93,Eckel10} It exhibits interesting transport phenomena such as Coulomb blockade and Kondo 
effect. The model comprises one spatial electronic state localized at the molecular bridge or the quantum dot (in the following referred to as bridge state), which can be occupied with two electrons of different spin polarization and two electronic continua describing the 
electrodes (leads). The Hamiltonian reads
\begin{eqnarray}\label{Htot}
        \hat H &=& \sum_{\sigma} E_d \hat{n}_{d,\sigma}
        + U_d \hat{n}_{d,\uparrow} \hat{n}_{d,\downarrow}
        + \sum_{k_L,\sigma} E_{k_L} \hat{n}_{k_L,\sigma}
        + \sum_{k_R,\sigma} E_{k_R} \hat{n}_{k_R,\sigma} \\
  &&    + \sum_{k_L,\sigma} V_{dk_L} ( \hat{d}^+_\sigma \hat{c}_{k_L,\sigma} 
        + \hat{c}_{k_L,\sigma}^+ \hat{d}_\sigma )
        + \sum_{k_R,\sigma} V_{dk_R} ( \hat{d}^+_\sigma \hat{c}_{k_R,\sigma} 
        + \hat{c}_{k_R,\sigma}^+ \hat{d}_\sigma ). \nonumber
\end{eqnarray}
In the above expression, $\hat{n}= \hat{d}^+ \hat{d}$ denotes the number operator, subscript ``$d$'' 
refers to the 
bridge state, ``$k_L/k_R$'' the states of the left/right metal leads, and 
``$\sigma=\uparrow,\downarrow$''
the two spin indices. Operators $\hat{d}^+/ \hat{d}$, $\hat{c}_{k_L}^+/ \hat{c}_{k_L}$, 
$\hat{c}_{k_R}^+/ \hat{c}_{k_R}$ are the fermionic creation/annihilation operators for the 
electronic states on the molecular bridge, the left and the right leads, respectively. 
The second term in Eq.\ (\ref{Htot}) describes
the on-site Coulomb repulsion of the electrons on the molecular bridge 
with electron-electron coupling strength $U_d$. The energies of the electronic
states in the leads, $E_{k_L}$, $E_{k_R}$, as well as the molecule-lead
coupling parameters  $V_{dk_L}$, $V_{dk_R}$ are assumed to be independent
of the spin polarization and are defined through the energy-dependent
level width functions
\begin{equation}
	\Gamma_L (E) = 2\pi \sum_{k_L} |V_{dk_L}|^2 \delta(E-E_{k_L}), \hspace{1cm}
	\Gamma_R (E) = 2\pi \sum_{k_R} |V_{dk_R}|^2 \delta(E-E_{k_R}).
\end{equation}

The energies and couplings in the above model can be obtained in various ways. One approach 
is to perform electronic structure calculations to extract the model parameters.\cite{Benesch09} In this paper, we consider a model parameterization for the 
electronic states of the leads, where the lead spectral density is an approximation to
a square band,
\begin{equation}\label{sb}
        \Gamma (E) = \frac{\Gamma}{[1+e^{(E-E_c)/\delta}][1+e^{-(E+E_c)/\delta}]}.
\end{equation}
In the limit $\delta\rightarrow 0$, $E = \pm E_c$ defines the band edge, with $\Gamma (E) = \Gamma$
in between.
The width functions for the left and the right leads are obtained by shifting $\Gamma(E)$ relative 
to the chemical potentials of the corresponding leads
\begin{equation}
	\Gamma_L (E) = \Gamma (E-\mu_L), \hspace{1cm}  \Gamma_R (E) = \Gamma (E-\mu_R),
\end{equation}
We consider a model of two identical leads, in which the chemical potentials are given by 
\begin{equation}
	\mu_{L/R} = E_f \pm V/2,
\end{equation}
where $V$ is the bias voltage and $E_f$ the Fermi energy of the leads. 

An observable X of interest for studying time-dependent transport can be expressed by the 
following time correlation function (in this paper we use atomic units where 
$\hbar = e = 1$)
\begin{equation}
	X(t) = \frac{1}{{\rm tr}[\hat{\rho}]} {\rm tr}
        \left[ \hat{\rho} \; e^{i\hat{H}t} \hat{X} e^{-i\hat{H}t} \right].
\end{equation}
For example, for a given bias voltage the time-dependent currents for the left 
and right leads are given by 
\begin{subequations}\label{sidecurrent}
\begin{equation}
	I_L(t) = - \frac{d N_L(t)} {dt} = -\frac{1}{{\rm tr}[\hat{\rho}]} {\rm tr}
        \left\{ \hat{\rho}\; e^{i\hat{H}t} i[\hat{H}, \hat{N}_{L}] e^{-i\hat{H}t} \right\},
\end{equation}
\begin{equation}
	I_R(t) = \frac{d N_R(t)} {dt} = \frac{1}{{\rm tr}[\hat{\rho}]} {\rm tr}
        \left\{ \hat{\rho}\; e^{i\hat{H}t} i[\hat{H}, \hat{N}_{R}] e^{-i\hat{H}t} \right\},
\end{equation}
\end{subequations}
where $N_{L/R}(t)$ denotes the  time-dependent charge in each lead 
\begin{equation}
	N_{\zeta}(t) = \frac{1}{{\rm tr}[\hat{\rho}]} {\rm tr}
        [\hat{\rho}\; e^{i\hat{H}t} \hat{N}_{\zeta} e^{-i\hat{H}t} ], \;\;\; \zeta=L, R,
\end{equation}
and  $\hat{N}_{\zeta} = \sum_{k_\zeta,\sigma} \hat{n}_{k_\zeta,\sigma}$
is the occupation number operator for the electrons in each lead ($\zeta=L, R$).

Other observables are also important for studying time-dependent quantum transport.  
Since the ML-MCTDH approach provides the wave function for the entire system, any properties can be obtained 
almost free of additional cost. For example, the electronic population of the bridge state 
with a particular spin $\sigma$ is given by
\begin{equation}
	n_{d,\sigma}(t) = \frac{1}{{\rm tr}[\hat{\rho}]} {\rm tr}
        [\hat{\rho}\; e^{i\hat{H}t} \hat{n}_{d,\zeta} e^{-i\hat{H}t} ],
\end{equation}
which can be used to analyze the nonequilibrium transport mechanisms.

\section{Details of the calculation}

In this section we discuss several important details of the calculation. Most of them are not 
only essential to the ML-MCTDH-SQR simulation but are also applicable to other approaches or 
have important physical relevance.

\subsection{The Initial Condition}

In the time correlation functions introduced in the previous section, $\hat{\rho}$ denotes the
initial density matrix. It is usually written in a factorized form (but see below for a 
different choice), i.e.,  it is modeled by a grand-canonical ensemble for each lead and 
a particular occupation of the bridge state,
\begin{subequations}\label{Initden}
\begin{equation}
        \hat{\rho}_e = \hat{\rho}_d^0 \;{\rm exp} \left[ -\beta (\hat{H}_e^0
          - \mu_L \hat{N}_L - \mu_R \hat{N}_R) \right],
\end{equation}
\begin{equation}
        \hat{H}_e^0 = \sum_{k_L,\sigma} E_{k_L} \hat{n}_{k_L,\sigma}
        + \sum_{k_R,\sigma} E_{k_R} \hat{n}_{k_R,\sigma}.
\end{equation}
\end{subequations}
Thereby, $\hat{\rho}_d^0$ is the initial reduced density matrix for the molecular bridge, which can be 
chosen as a pure state representing either occupied or empty bridge states. Other initial states 
may also be used. For example, one may use a fully correlated initial state,
\begin{equation}\label{corrden}
        \hat{\rho} = {\rm exp} \left[ -\beta (\hat{H}
          - \mu_L \hat{N}_L - \mu_R \hat{N}_R) \right],
\end{equation}
employing the imaginary time ML-MCTDH propagation, which is similar to what we have proposed previously for a canonical
ensemble.\cite{wan06:034114}  This is particularly advantageous for studying the steady-state current at strong 
coupling regimes where a factorized initial state results in strong transient effect in the current.

\subsection{Discretization of the Nuclear and Electronic Continua}

Since ML-MCTDH simulations employ wave functions explicitly for all degrees of freedom, the 
continua of the two electrodes need to be discretized. This is similar to a numerical integration, 
and there are various ways to achieve the goal of an efficient discretization. Formally, the electronic continuum for the two 
leads can be discretized by choosing a density of states $\rho_e(E)$ such 
that\cite{tho01:2991,wan01:2979,wan03:1289}
\begin{equation}\label{olddis}
	\int_{-E_M}^{E_k} dE \; \rho_e(E) = k, \hspace{.5in} 
	|V_{dk}|^2 =  \frac{\Gamma(E_k)}{2\pi \rho_e(E_k)}, \hspace{.5in}
	 k = 1,...,N_e,
\end{equation}
where $N_e$ is the number of electronic states (for a single spin/single lead) in the simulation.
This is essentially a quadrature representation of an integral that involves $\Gamma(E)$.  Very
often a constant $\rho_e(E)$, i.e., an equidistant discretization of the whole band interval, works
best for representing the electronic continuum.  This corresponds to a sinc-function quadrature 
which is known to be effective for handling oscillatory integrands.  However, for some problems 
it is well-known (and we will show this numerically in this paper) that Wilson's logarithmic 
discretization scheme\cite{Wilson75} is more efficient to focus on the electronic states near the Fermi level.  
For states above the Fermi level, this discretization scheme gives
\begin{equation}\label{log}
	\frac{E_k - E_f}{E_M - E_k} = \Lambda^{M - k},
\end{equation}
where $\Lambda \rightarrow 1^+$, $M = N_e/2$, and $E_M$ is the energy of the band edge.  For a
symmetric band, the states below the Fermi level are obtained by symmetry, otherwise the above 
relation is simply modified via $E \rightarrow -E$.   

From a general perspective, Wilson's logarithmic discretization employs a relation similar to (\ref{olddis}) 
based on a particular density of states
\begin{equation}
	\int_{E_k}^{E_M} dE \; \rho_e(E) = M - k, 
\end{equation}
It is easy to verify that the density of states satisfying Eq.~(\ref{log}) is given by
\begin{equation}\label{whole}
	\rho_e(E) = \frac{1}{(E - E_f){\rm ln} \Lambda}. 
\end{equation}
In many applications, one does not take $E_k$ in Eq.~(\ref{log}) but rather an average energy $\bar{E}_k$,
\begin{subequations}
\begin{equation}
	\bar{E}_k = \frac{\int^k dE \; \rho_e(E) E}{\int^k dE \; \rho_e(E)},
\end{equation}
where
\begin{equation}
	\int^k dE \equiv \int_{E_{k-1}}^{E_k} dE.
\end{equation}
\end{subequations}
Moreover, the density of states $\rho_e(E)$ does not need to be a continuous function in the whole interval
as Eq.~(\ref{whole}).  Instead, in many NRG applications $\rho_e(E)$ is chosen to be piecewise continuous in
each of the $[E_{k-1}, E_k]$ interval, where $E_k$ is given in (\ref{log}).  Within each such interval one
is free to choose a particular $\rho_e(E)$ such that
\begin{equation}\label{disrho}
	\int^k dE \; \rho_e(E)  = 1.
\end{equation}
For example, one may require that $\rho_e(E) \propto \Gamma(E)$ in each interval as done in many NRG applications.
Our investigation shows that these different variants make negligible difference in our applications when the 
number of discrete states is large enough (e.g., $>$ 100).

\subsection{Regularization of the Current}

The transient behavior of the single-lead currents $I_R(t)$ and $I_L(t)$ defined in Eq.~(\ref{sidecurrent}) 
is usually different. However, the long-time limits of $I_R(t)$ and $I_L(t)$, which define the steady-state 
current, are the same. It is found that the average current 
\begin{equation}
	I(t) = \frac{1}{2} [ I_R(t) + I_L(t) ],
\end{equation} 
provides better numerical convergence properties by minimizing the transient characteristic, and 
thus will be used in most calculations.

Since the electronic continuum is represented by a finite number of states, recurrences will eventually occur at longer times.  
This time scale depends on the number of states used in the simulation. The situation is thus similar to that 
of a quantum reactive scattering calculation in the presence of a scattering continuum, where, with a finite 
number of basis functions, an appropriate absorbing boundary condition is added to mimic the correct outgoing 
Green's function.\cite{Goldberg1978,Kosloff1986,Neuhauser1989,Seideman1991} 
Employing the same strategy for the present problem, the 
regularized electric current is given by
\begin{equation}
	I^{\rm reg}  = \lim_{\eta \to 0^+} \int_0^{\infty} dt \, \frac{dI(t)}{dt} \, e^{-\eta t}.
\end{equation}
The regularization parameter $\eta$ is similar (though not identical) to the formal convergence parameter
in the definition of  the Green's function in terms of the time evolution operator
\begin{equation}
	G(E^+) = \lim_{\eta \to 0^+} (-i) \int_0^{\infty} dt\, e^{i(E+i\eta-H)t}.
\end{equation}
In numerical calculations, the parameter $\eta$ has to be large enough to accelerate the convergence but 
still sufficiently small in order not to affect the correct result.  While in the reactive scattering calculation
$\eta$ is often chosen to be coordinate dependent, here $\eta$ is chosen to be time dependent
\begin{equation}\label{damping}
\eta(t) = \left\{
              \begin{array}{ll}
                   0 & \quad (t<\tau)\\
                   \eta_0\cdot (t-\tau)/t & \quad (t>\tau) .
              \end{array}
       \right.
\end{equation}
Here $\eta_0$ is a damping constant, $\tau$ is a cutoff time beyond which a steady state charge
flow is approximately reached.  As the number of electronic states increases, one may choose a 
weaker damping strength $\eta_0$ and/or longer cutoff time $\tau$.  The former approaches zero and the
latter approaches infinity for an infinite number of states.

\section{The Multilayer Multiconfiguration Time-Dependent Hartree Theory in Second Quantization Representation}\label{mlsqr}

To describe many-body quantum dynamics in an efficient way, we employ the multilayer 
multiconfiguration time-dependent Hartree (ML-MCTDH) theory in second quantization 
representation (SQR).\cite{wan10:78} The ML-MCTDH theory\cite{wan03:1289,Wang15} is a rigorous 
variational method to propagate wave packets in complex systems with many degrees of freedom.  
In this approach the wave function is represented by a recursive, layered expansion, 
\begin{subequations}\label{psiml}
\begin{equation}\label{L1}
        |\Psi (t) \rangle = \sum_{j_1} \sum_{j_2} ... \sum_{j_p}
        A_{j_1j_2...j_p}(t) \prod_{\kappa=1}^{p}  |\varphi_{j_\kappa}^{(\kappa)} (t) \rangle,
\end{equation}
\begin{equation}\label{L2}
        |\varphi_{j_\kappa}^{(\kappa)}(t)\rangle =  \sum_{i_1} \sum_{i_2} ... \sum_{i_{Q(\kappa)}}
        B_{i_1i_2...i_{Q(\kappa)}}^{\kappa,j_\kappa}(t) \prod_{q=1}^{Q(\kappa)}  
	|v_{i_q}^{(\kappa,q)}(t) \rangle,
\end{equation}
\begin{equation}\label{L3}
        |v_{i_q}^{(\kappa,q)}(t)\rangle  = \sum_{\alpha_1} \sum_{\alpha_2} ... 
	\sum_{\alpha_{M(\kappa,q)}}
        C_{\alpha_1\alpha_2...\alpha_{M(\kappa,q)}}^{\kappa,q,i_q}(t) 
	\prod_{\gamma=1}^{M(\kappa,q)}  
	|\xi_{\alpha_\gamma}^{\kappa,q,\gamma}(t) \rangle,
\end{equation}
\begin{equation}
	... \nonumber
\end{equation}
\end{subequations}
where $A_{j_1j_2...j_p}(t)$, $B_{i_1i_2...i_{Q(\kappa)}}^{\kappa,j_\kappa}(t)$,
$C_{\alpha_1\alpha_2...\alpha_{M(\kappa,q)}}^{\kappa,q,i_q}(t)$ and so on are the
expansion coefficients for the first, second, third, ..., layers, respectively;
$|\varphi_{j_\kappa}^{(\kappa)} (t) \rangle$, $|v_{i_q}^{(\kappa,q)}(t) \rangle$,
$|\xi_{\alpha_\gamma}^{\kappa,q,\gamma}(t) \rangle$, ..., are the ``single particle'' 
functions (SPFs) for the first, second, third, ..., layers.  
In Eq.~(\ref{L1}), $p$ denotes the number of single
particle (SP) groups/subspaces for the first layer.  Similarly, $Q(\kappa)$ in Eq.~(\ref{L2})
is the number of SP groups for the second layer that belongs to the $\kappa$th SP
group in the first layer,  i.e., there are a total of $\sum_{\kappa=1}^{p} Q(\kappa)$
second layer SP groups.  Continuing along the multilayer hierarchy, 
$M(\kappa,q)$ in Eq.~(\ref{L3}) is the number of SP groups for the third layer that belongs 
to the $q$th SP group of the second layer and the $\kappa$th SP group of the first layer,  
resulting in a total of $\sum_{\kappa=1}^{p} \sum_{q=1}^{Q(\kappa)} M(\kappa,q)$ third 
layer SP groups.  

ML-MCTDH is an effective tensor contraction scheme. Its mathematical form in (\ref{psiml}) 
has been given various names such as the hierarchical low rank tensor format, the tree Tucker 
format, the tensor train format, and the sequential unfolding SVD.\cite{Wang15} The size 
of the system that the ML-MCTDH theory can treat increases with the number of layers in the 
expansion.  In principle, such a recursive expansion can be carried out to an arbitrary number 
of layers.  The multilayer hierarchy is terminated at a particular level by expanding the SPFs 
in the deepest layer in terms of time-independent configurations, each of which may still contain 
several Cartesian degrees of freedom.
The variational parameters within the ML-MCTDH theoretical framework are dynamically 
optimized through the use of the Dirac-Frenkel variational principle\cite{Frenkel34}
\begin{equation}
        \langle \delta\Psi(t) | i \frac{\partial} {\partial t} - \hat{H} |
        \Psi(t) \rangle = 0,
\end{equation}
which results in a set of coupled, nonlinear differential equations for the expansion
coefficients for all layers.\cite{wan03:1289,wan10:78,wan09:024114} 

The development of the ML-MCTDH method was motivated by the original MCTDH 
method\cite{mey90:73,man92:3199,bec00:1,mey03:251,Meyer09} and extends its applicability to significantly 
larger systems.\cite{wan03:1289,Thoss04b,tho06:210,wan06:034114,wan07:10369,cra07:144503} 
The theory has also been generalized to study heat transport in molecular 
junctions\cite{vel08:325} and to calculate thermal rate constants for condensed phase
systems using the reactive flux correlation function formalism.\cite{wan06:174502,cra07:144503}
The work of Manthe has introduced an even more adaptive formulation based on
a layered correlation discrete variable representation (CDVR).\cite{man08:164116,man09:054109}
ML-MCTDH has also been implemented in the popular Heidelberg MCTDH program package.\cite{Meyer09}  
Recently, the ML-MCTDH theory has been formulated within an interaction picture to remove artificial correlation.\cite{Wang16,Wang17}

Another hurdle to overcome for the study of electron transport processes is that the methodology
has to be adapted to handle identical quantum particles, i.e., to incorporate the exchange symmetry 
explicitly. Within the (single-layer) MCTDH method, one may employ a properly symmetrized wave 
function in the first quantized framework, i.e., permanents in a bosonic case\cite{alo08:033613} or 
Slater determinants in a fermionic case.\cite{kat04:533,cai05:012712,nes05:124102} However, this wave 
function-based symmetrization is incompatible with the ML-MCTDH theory with more layers --- there is 
no obvious analog of a multilayer Hartree configuration if permanents/determinants are used to 
represent the wave function. Therefore, we proposed a novel approach, the ML-MCTDH-SQR method,\cite{wan09:024114} that follows 
a fundamentally different route to tackle many-body quantum dynamics of indistinguishable particles 
--- an operator-based method that employs the second quantization formalism of many-particle quantum 
theory.  This differs from many previous methods where the second quantization formalism is only used 
as a convenient tool to derive intermediate expressions for the first quantized form.  In the ML-MCTDH-SQR 
approach the variation is carried out entirely in the abstract Fock space using the occupation number 
representation. Therefore, the burden of handling symmetries of identical particles in a numerical 
variational calculation is shifted completely from wave functions to the algebraic properties of 
operators.  For example, in the second quantized form the fermionic creation/annihilation operators 
fulfill the anti-commutation relations
\begin{equation}\label{anticomm}
	\{ \hat{a}_P, \hat{a}_Q^+ \} \equiv \hat{a}_P \hat{a}_Q^+ + \hat{a}_Q^+ \hat{a}_P = \delta_{PQ},
	\hspace{1cm}
	\{ \hat{a}_P^+, \hat{a}_Q^+ \} = \{ \hat{a}_P, \hat{a}_Q \} = 0.
\end{equation}
The symmetry of identical particles is thus realized by enforcing such algebraic
properties of the operators.  This can be accomplished by introducing a permutation
sign operator associated with each fermionic creation/annihilation operator, which incorporates 
the sign changes of the remaining spin orbitals in all the SPFs whose subspaces are prior to 
it.\cite{wan09:024114} 

In the second quantized form, the wave function is represented in the abstract Fock space 
employing the occupation number basis.  This is not plausible if one is only interested in a
single configuration-based method, e.g., Hartree-Fock or Kohn-Sham density functional theory.
However, it is advantageous when one aims at treating correlation effect in a numerically exact
way via a multiconfigurational method, especially in the multilayer form.  Within the second
quantization representation, any wave function can be easily expanded in the same multilayer 
form as that for systems of distinguishable particles.  All the ML-MCTDH techniques can thus
be adopted naturally. The symmetry of the wave function in the first quantized form is shifted 
to the operator algebra in the second quantized form.  
The key point is that, for both phenomenological models and more fundamental theories, 
there are only a limited number of combination of fundamental operators.  
For example, in electronic structure theory only one- and two-electron operators are present.  
This means that one never needs to handle all, redundant possibilities of operator combinations as 
offered by the determinant form in the first quantized framework.  It is exactly this property 
that provides the flexibility of representing the wave functions in multilayer form and treat 
them accurately and efficiently within the ML-MCTDH theory.

It is noted that Manthe et al.\ have presented a variant of the ML-MCTDH-SQR approach using optimized time-dependent orbitals.\cite{Manthe17} Furthermore, the concept of the SQR has also been used recently within a single layer MCTDH implementation to study impurity models.\cite{Balzer15}

\section{Results and Discussion}\label{results}

In this section, we apply the ML-MCTDH methodology outlined above to the Anderson impurity model at zero 
temperature. In this regime, the Anderson impurity model is known to exhibit Kondo effect,\cite{Hewson93,Wiel00,lia02:725} which manifests itself in a many-body resonance of the conductance at the Fermi energy. We specifically consider the particle-hole symmetric regime with $E_d = - U_d/2$, where the Kondo resonance for the zero-bias conductance exhibits a maximal value of the conductance quantum $G_0$.  

\begin{figure}[!ht]
\includegraphics[clip,width=0.45\textwidth]{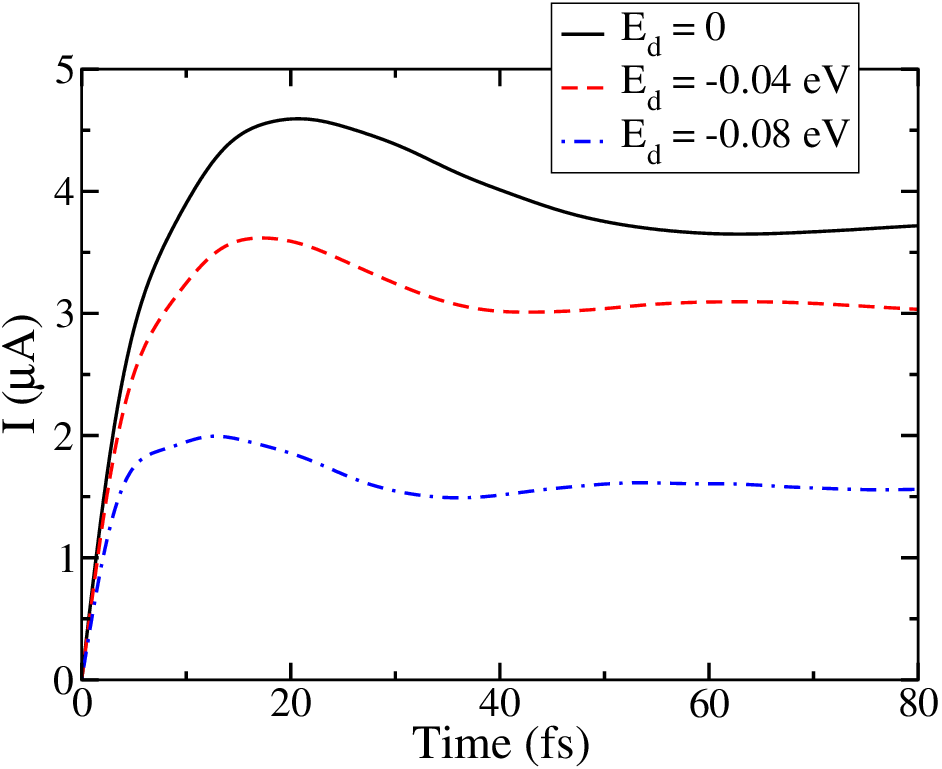}

\caption{Time-dependent current for bias voltage $V = 0.1V$ and zero temperature.  The square model of Eq.~(\ref{sb}) 
is used. The parameters are: $\delta = 4\times 10^{-4}$ eV, $\Gamma_{L/R} = 0.02$ eV, $E_c = 0.4$ eV, $E_f = 0$,
and maintaining $E_d = - U_d/2$. }
\label{V_0.1}
\end{figure}

Figure~\ref{V_0.1} shows the current as a function of time for a bias voltage of $V = 0.1$ V and different values of Coulomb repulsion $U_d = -2 E_d$. After a transient regime the current saturates to a steady state plateau, the value of which decreases for larger $U_d$, in accordance with the energy level scheme.

For the voltage of $V = 0.1$ eV used in Figure~\ref{V_0.1}, which is already outside the linear response regime, the current can be simulated directly as discussed above. For smaller voltages, as is necessary to access, e.g., the zero-bias conductance, the direct simulation of the current  using the standard factorized initial density matrix and equal-spaced discretization of the electronic continuum becomes numerically challenging.  This 
can be attributed to two physical reasons.  First, due to the large transient peaks in the time-dependent 
current, the noise to signal ratio for the steady state current becomes large.  Second, to facilitate the 
current at very small bias voltage, states of the electrodes near the Fermi level become important.  
An equal-spaced discretization is in this case inefficient and the calculation would require many more states 
to converge.

The second issue is related to the particular physics in the Kondo regime and has been studied previously. 
To focus on the states near the Fermi level, the scheme pinoneered by Wilson,\cite{Wilson75} 
Eqs.~(\ref{log})-(\ref{disrho}), has proven to be very effective.  It is widely used in numerical 
renormalization group (NRG) theory calculations, typically with a relatively large value of the parameter  $\Lambda \sim 2$.  Unlike in NRG, 
the ML-MCTDH simulation can easily treat more than 200 states per lead, which enables us to examine 
the limit $\Lambda \rightarrow 1^+$ in Eq.~(\ref{log}).

\begin{figure}[!ht]
\includegraphics[clip,width=0.45\textwidth]{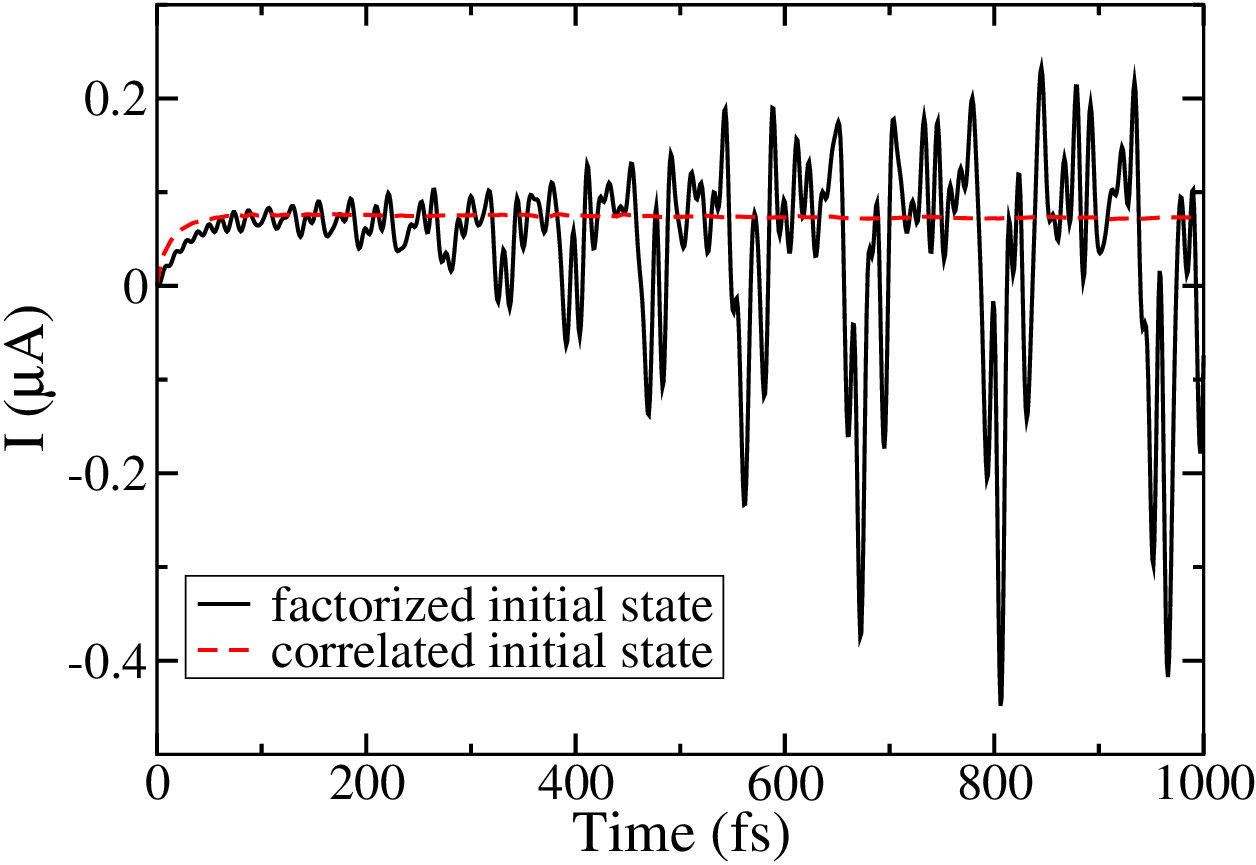}

\caption{Time-dependent current for bias voltage $V = 0.001V$ and zero temperature.  The square model of Eq.~(\ref{sb}) 
is used, where 216 discrete states per lead are employed in the logrithmic discretization, Eq.~(\ref{log}),
with $\Lambda = 1.175$. The parameters are the same as in Fig.~\ref{V_0.1} except $E_d = U_d = 0$.  For the
correlated initial state a fictitious temperature of $T=2.5 K$ was used in the imaginary-time relaxation calculation. }
\label{V_0.001}
\end{figure}

The use of Wilson's logarithmic discretization scheme, however, does not solve the first problem related 
to the factorized initial density matrix. In fact, it makes the situation even worse since the electronic 
states are no longer arranged as Fourier grids. As a result, the time-dependent current exhibits 
severe oscillations, which is demonstrated in Figure~\ref{V_0.001}. This problem is also known in NRG, where 
a time-averaging scheme is often employed to smooth the long-time current.\cite{and08:066804} For this purpuse one may also 
apply the regularization scheme described earlier in this paper.

A more robust way towards the same goal is to employ a correlated initial density matrix, 
Eq.~(\ref{corrden}).  At a finite temperature this is achieved by combining imaginary time ML-MCTDH 
propagation with a Monte Carlo importance sampling scheme.\cite{wan06:034114}  For the case of zero 
temperture, it is done as an imaginary time ML-MCTDH propagation (relaxation) with a sufficiently 
large imaginary time $\beta=1/k_BT$. As shown in Figure~\ref{V_0.001}, the use of a correlated initial 
density matrix improves the situation significantly. The current fluctuates only within a few percent 
relative deviation, which does not show up in the scale of the plot. Within the numerical
resolution, $T=2.5 K$ for the imaginary time propagation is satisfactory for this example. Nevertheless,
this is a convergence parameter and needs to be tested.  For the parameter regime discussed in this
paper convergence is achieved within the range of 0.5 - 10 K. 

\begin{figure}[!ht]
\includegraphics[clip,width=0.45\textwidth]{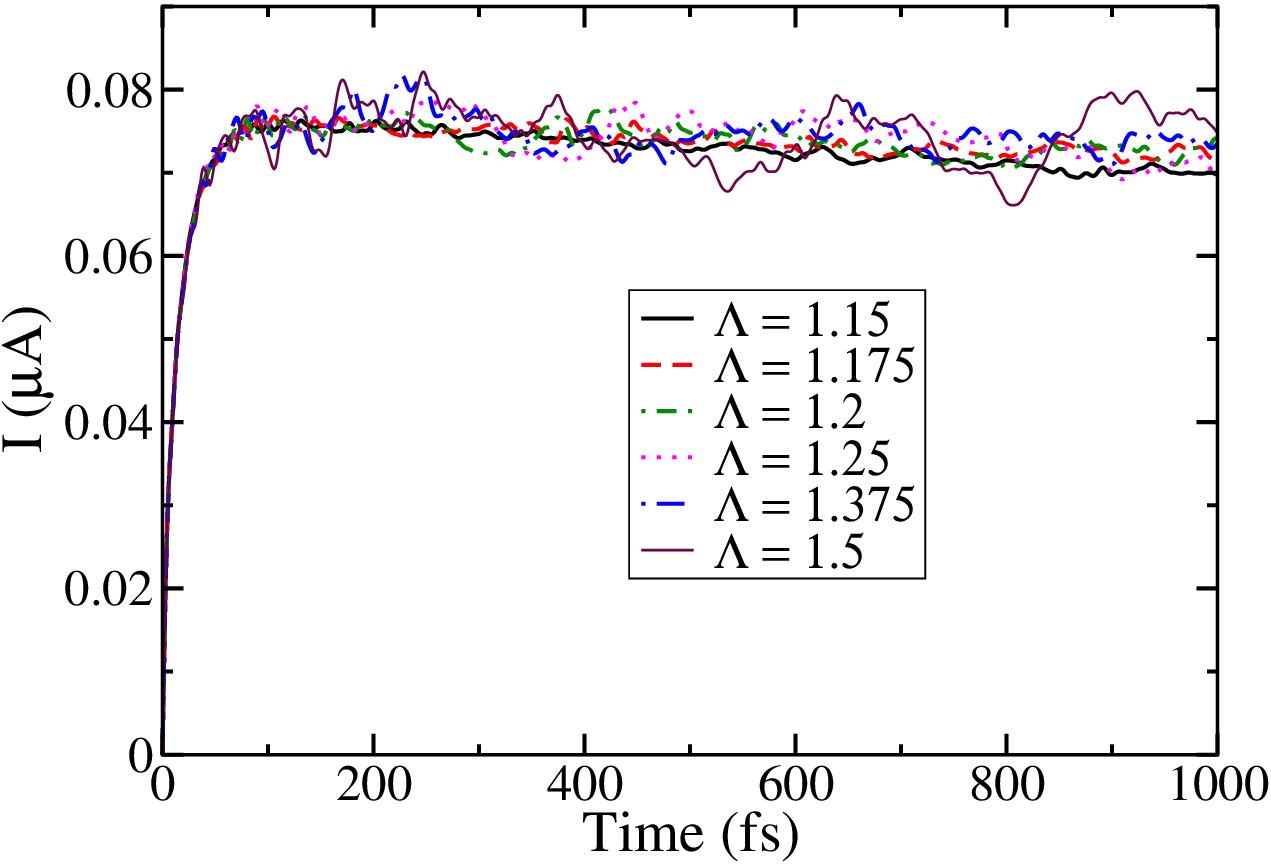}

\caption{Time-dependent current vesus parameter $\Lambda$ in Eq.~(\ref{log}). The parameters are the same 
as in Fig.~\ref{V_0.001}, where a correlated initial state is prepared using imaginary time ML-MCTDH propagation with 
a fictitious temperature of $T=2.5 K$. }
\label{Lambda_change}
\end{figure}

Within Wilson's logarithmic discretization scheme, varying the parameter $\Lambda$ in Eq.~(\ref{log}) has some influence on the value of the calculated current, as shown in  Figure~\ref{Lambda_change}. For a fixed 
number of discrete states, a too small $\Lambda$ results in an insufficient number of states near the 
Fermi level and thus too small a current.  On the other hand, a larger $\Lambda$ results in a larger 
amplitude of oscillation in the current.  Convergence is achieved by reducing $\Lambda$ while 
increasing the number of discrete states at the same time.  In Figure~\ref{Lambda_change}, most steady-state 
currents agree well with each other except for the smallest and the largest $\Lambda$. 

\begin{figure}[!ht]
\includegraphics[clip,width=0.45\textwidth]{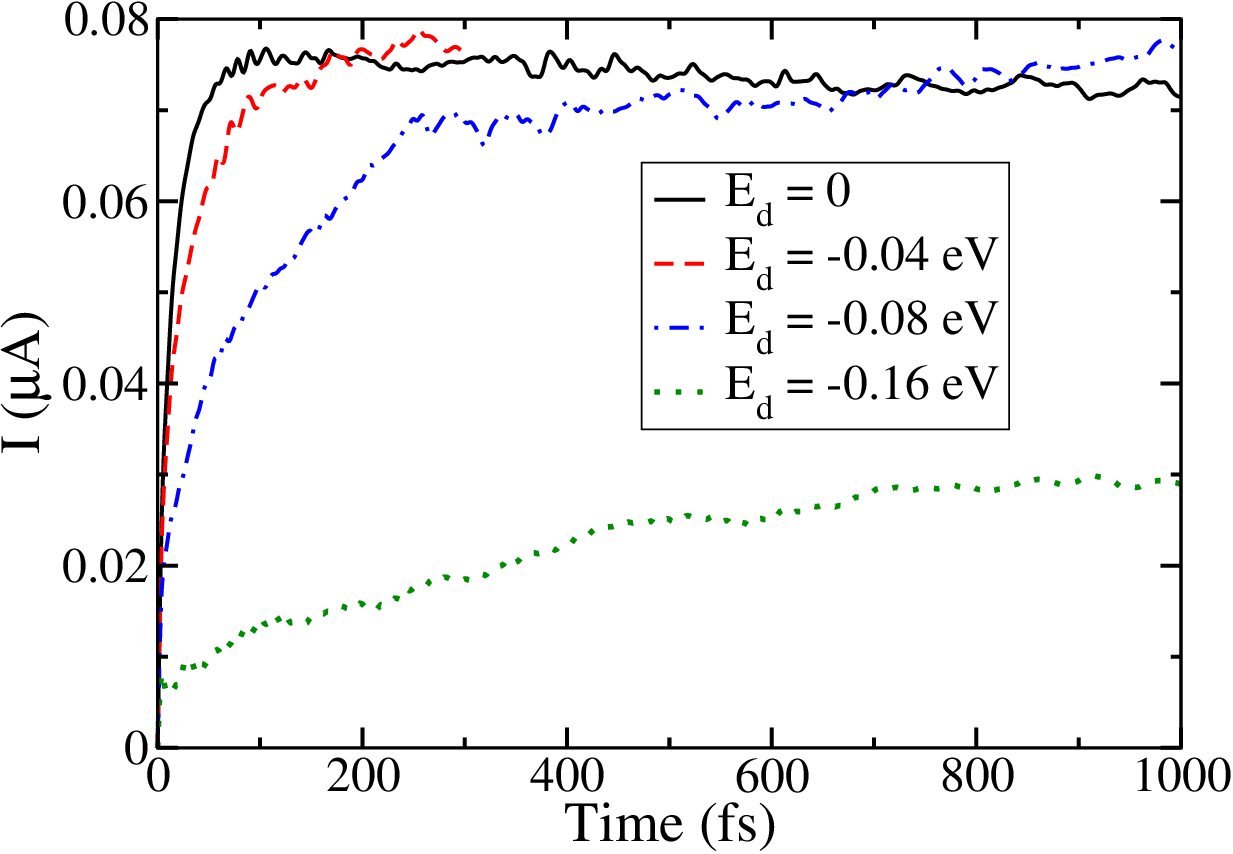}

\caption{Time-dependent current for bias voltage $V = 0.001V$ and zero temperature.  
The square model of Eq.~(\ref{sb}) is used. The parameters are  the same as in Fig.~\ref{Lambda_change}, 
while keeping $E_d = - U_d/2$. }
\label{V_0.001_Kondo}
\end{figure}

With these methodological improvements, it is possible to examine the steady-state current in the Kondo 
regime even for very small voltages.  As an example, Figure~\ref{V_0.001_Kondo} shows the time-dependent current for a bias voltage of 0.001V. 
For not too large Coulomb repulsion $U_d$, the steady-state current (and hence the conductance) is the 
same for different values of $E_d$ as long as $E_d = - U_d/2$ is maintained.  For larger values, e.g.\ $E_d = - U_d/2 = -0.16$ eV, the conductance is however no longer in the linear regime and therefore attains a smaller value. Simulations 
for even smaller bias voltage $V=0.0001V$ show that $E_d = - U_d/2 = -0.16$ eV has the same steady-state 
current as for the other values of $E_d$ and, thus, is in the linear conductance regime.  

\begin{figure}[!ht]
\begin{flushleft}
(a)
\end{flushleft}
\includegraphics[clip,width=0.45\textwidth]{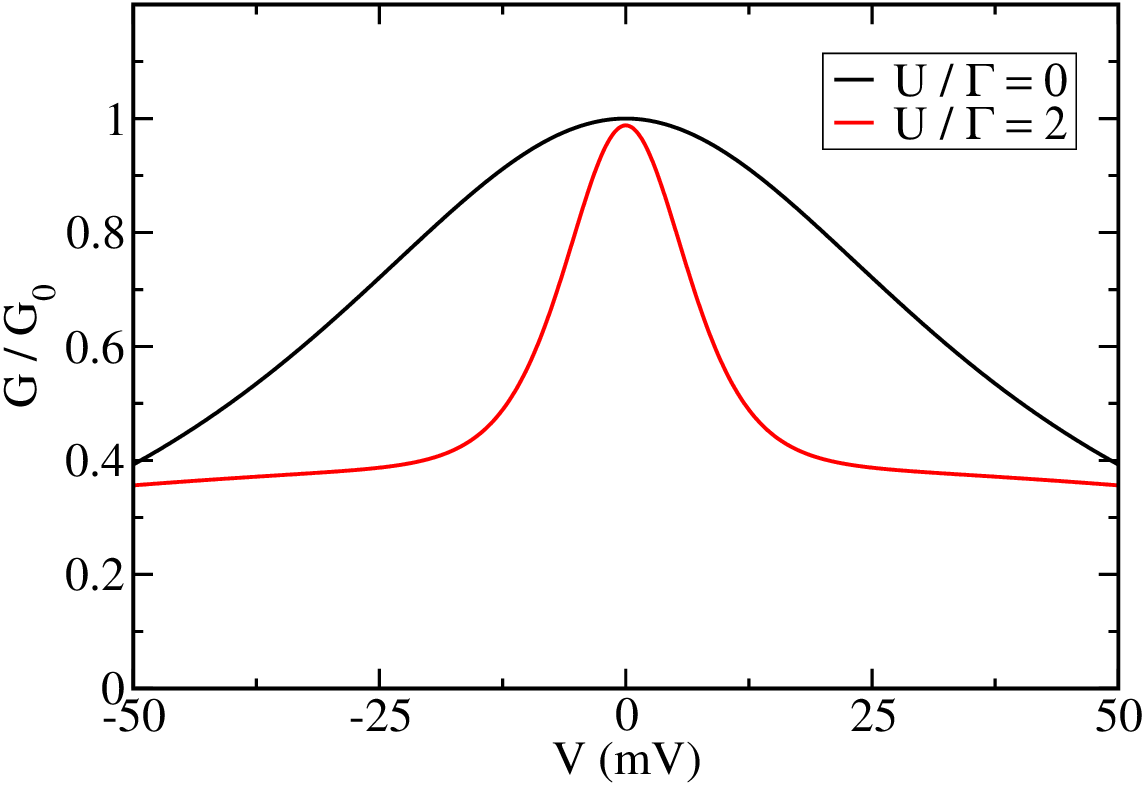}

\begin{flushleft}
(b)
\end{flushleft}
\includegraphics[clip,width=0.45\textwidth]{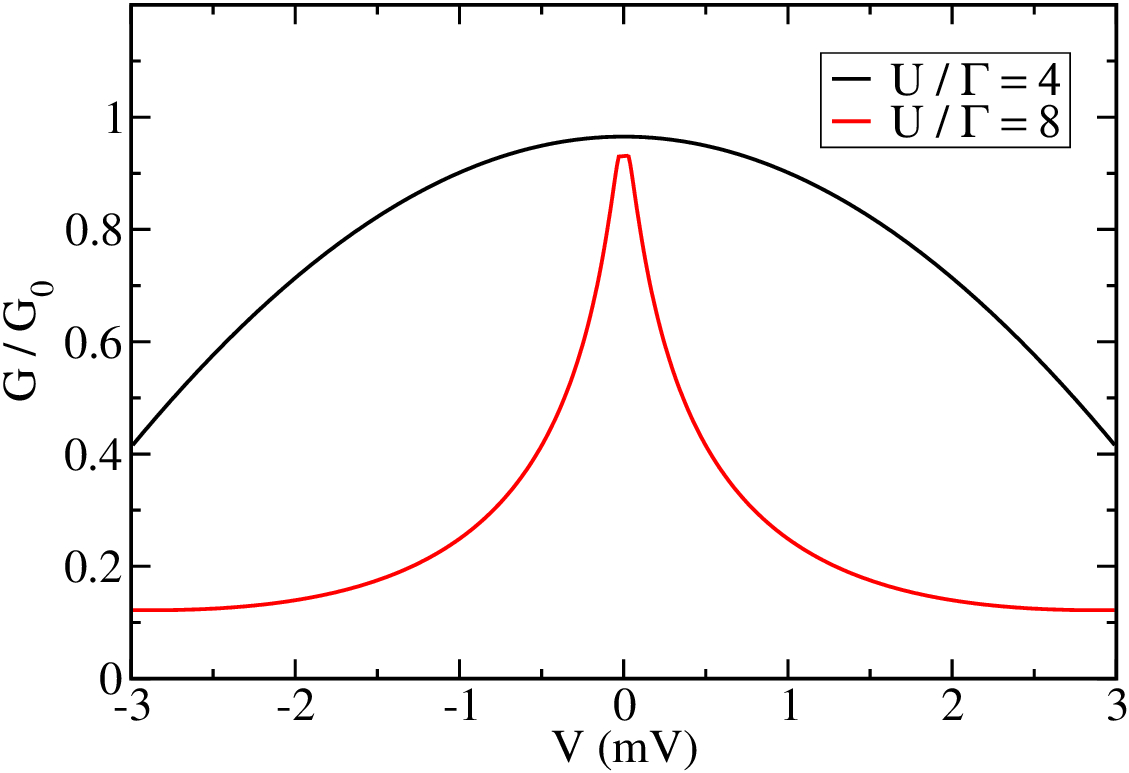}

\caption{Conductance versus bias voltage. The square model of Eq.~(\ref{sb}) 
is used. The parameters are the same as in Fig.~\ref{Lambda_change}, while keeping $E_d = - U_d/2$. }
\label{Cond_Kondo}
\end{figure}

Figure~\ref{Cond_Kondo} shows the conductance as a function of the bias voltage for different Coulomb repulsion strengths, where the relation $E_d = - U_d/2$ is maintained.
The transition from the broad single-particle resonance for the noninteracting model ($U_d = 0$) to the many-body Kondo resonance is seen. As the electron-electron coupling increases, the width of the Kondo resonance becomes narrower, in qualitative accordance with approximate analytical theories.\cite{Oguri05}

\section{Concluding Remarks}\label{conclusions}

In this paper, we have employed the ML-MCTDH-SQR method to study correlated electron transport 
in the Anderson-impurity model at zero temperature. Extending our previous work,\cite{wan09:024114,Wang13} we have implemented
Wilson's logarithmic discretization scheme for the electronic continuum, such that states near 
the Fermi level are taken into account with a higher weight, and the imaginary time propagation to obtain a correlated
initial wave function. These improvements of  the methodology allow us to address both linear and nonlinear conductance and to study more difficult physical regimes.

We have specifically focused on nonequilibrium transport at zero temperature. In this regime, the Anderson impurity model exhibits Kondo effect. Our calculations reveal the Kondo resonance and show the dependence of the Kondo peak line shape on the strength of the Coulomb repulsion. The ability of obtaining this limit demonstrates that ML-MCTDH-SQR is a useful theory for studying nonequilibrium correlated quantum transport.

\section*{Acknowledgments}
This paper is dedicated to Hans-Dieter Meyer on the occasion of his 70th birthday.
The work has been supported by the National Science Foundation CHE-1500285 (HW) and the 
German Science Foundation (DFG) (MT) through SFB 953 and a research grant, and used resources of the National Energy 
Research Scientific Computing Center (NERSC), which is supported by the Office of Science of the U.S.  Department of Energy under Contract No. DE-AC02-05CH11231.

\pagebreak

\clearpage
~\vspace{3cm}

\clearpage
~\vspace{3cm}

\clearpage
~\vspace{3cm}

\clearpage
~\vspace{3cm}

\clearpage
~\vspace{1cm}

\begin{thebibliography}{100}
\expandafter\ifx\csname bibnamefont\endcsname\relax
  \def\bibnamefont#1{#1}\fi
\expandafter\ifx\csname bibfnamefont\endcsname\relax
  \def\bibfnamefont#1{#1}\fi
\expandafter\ifx\csname url\endcsname\relax
  \def\url#1{\texttt{#1}}\fi
\expandafter\ifx\csname urlprefix\endcsname\relax\def\urlprefix{URL }\fi
\providecommand{\bibinfo}[2]{#2}
\providecommand{\eprint}[2][]{\url{#2}}

\bibitem{ree97:252}
\bibinfo{author}{\bibfnamefont{M.}~\bibnamefont{Reed}},
  \bibinfo{author}{\bibfnamefont{C.}~\bibnamefont{Zhou}},
  \bibinfo{author}{\bibfnamefont{C.}~\bibnamefont{Muller}},
  \bibinfo{author}{\bibfnamefont{T.}~\bibnamefont{Burgin}}, \bibnamefont{and}
  \bibinfo{author}{\bibfnamefont{J.}~\bibnamefont{Tour}},
  \bibinfo{journal}{Science} \textbf{\bibinfo{volume}{278}},
  \bibinfo{pages}{252} (\bibinfo{year}{1997}).

\bibitem{joa00:541}
\bibinfo{author}{\bibfnamefont{C.}~\bibnamefont{Joachim}},
  \bibinfo{author}{\bibfnamefont{J.}~\bibnamefont{Gimzewski}},
  \bibnamefont{and} \bibinfo{author}{\bibfnamefont{A.}~\bibnamefont{Aviram}},
  \bibinfo{journal}{Nature (London)} \textbf{\bibinfo{volume}{408}},
  \bibinfo{pages}{541} (\bibinfo{year}{2000}).

\bibitem{Nitzan01}
\bibinfo{author}{\bibfnamefont{A.}~\bibnamefont{Nitzan}},
  \bibinfo{journal}{Annu. Rev. Phys. Chem.} \textbf{\bibinfo{volume}{52}},
  \bibinfo{pages}{681} (\bibinfo{year}{2001}).

\bibitem{nit03:1384}
\bibinfo{author}{\bibfnamefont{A.}~\bibnamefont{Nitzan}} \bibnamefont{and}
  \bibinfo{author}{\bibfnamefont{M.~A.} \bibnamefont{Ratner}},
  \bibinfo{journal}{Science} \textbf{\bibinfo{volume}{300}},
  \bibinfo{pages}{1384} (\bibinfo{year}{2003}).

\bibitem{Cuniberti05}
\bibinfo{author}{\bibfnamefont{G.}~\bibnamefont{Cuniberti}},
  \bibinfo{author}{\bibfnamefont{G.}~\bibnamefont{Fagas}}, \bibnamefont{and}
  \bibinfo{author}{\bibfnamefont{K.}~\bibnamefont{Richter}},
  \emph{\bibinfo{title}{Introducing Molecular Electronics}}
  (\bibinfo{publisher}{Springer}, \bibinfo{address}{Heidelberg},
  \bibinfo{year}{2005}).

\bibitem{Selzer06}
\bibinfo{author}{\bibfnamefont{Y.}~\bibnamefont{Selzer}} \bibnamefont{and}
  \bibinfo{author}{\bibfnamefont{D.~L.} \bibnamefont{Allara}},
  \bibinfo{journal}{Annu. Rev. Phys. Chem.} \textbf{\bibinfo{volume}{57}},
  \bibinfo{pages}{593} (\bibinfo{year}{2006}).

\bibitem{Venkataraman06}
\bibinfo{author}{\bibfnamefont{L.}~\bibnamefont{Venkataraman}},
  \bibinfo{author}{\bibfnamefont{J.~E.} \bibnamefont{Klare}},
  \bibinfo{author}{\bibfnamefont{C.}~\bibnamefont{Nuckolls}},
  \bibinfo{author}{\bibfnamefont{M.~S.} \bibnamefont{Hybertsen}},
  \bibnamefont{and} \bibinfo{author}{\bibfnamefont{M.~L.}
  \bibnamefont{Steigerwald}}, \bibinfo{journal}{Nature}
  \textbf{\bibinfo{volume}{442}}, \bibinfo{pages}{904} (\bibinfo{year}{2006}).

\bibitem{Chen07}
\bibinfo{author}{\bibfnamefont{F.}~\bibnamefont{Chen}},
  \bibinfo{author}{\bibfnamefont{J.}~\bibnamefont{Hihath}},
  \bibinfo{author}{\bibfnamefont{Z.}~\bibnamefont{Huang}},
  \bibinfo{author}{\bibfnamefont{X.}~\bibnamefont{Li}}, \bibnamefont{and}
  \bibinfo{author}{\bibfnamefont{N.}~\bibnamefont{Tao}},
  \bibinfo{journal}{Annu. Rev. Phys. Chem.} \textbf{\bibinfo{volume}{58}},
  \bibinfo{pages}{535} (\bibinfo{year}{2007}).

\bibitem{Galperin08b}
\bibinfo{author}{\bibfnamefont{M.}~\bibnamefont{Galperin}},
  \bibinfo{author}{\bibfnamefont{M.~A.} \bibnamefont{Ratner}},
  \bibinfo{author}{\bibfnamefont{A.}~\bibnamefont{Nitzan}}, \bibnamefont{and}
  \bibinfo{author}{\bibfnamefont{A.}~\bibnamefont{Troisi}},
  \bibinfo{journal}{Science} \textbf{\bibinfo{volume}{319}},
  \bibinfo{pages}{1056} (\bibinfo{year}{2008}).

\bibitem{Cuevas10}
\bibinfo{author}{\bibfnamefont{J.}~\bibnamefont{Cuevas}} \bibnamefont{and}
  \bibinfo{author}{\bibfnamefont{E.}~\bibnamefont{Scheer}},
  \emph{\bibinfo{title}{Molecular Electronics: An Introduction to Theory and
  Experiment}} (\bibinfo{publisher}{World Scientific},
  \bibinfo{address}{Singapore}, \bibinfo{year}{2010}).

\bibitem{par02:722}
\bibinfo{author}{\bibfnamefont{J.}~\bibnamefont{Park}},
  \bibinfo{author}{\bibfnamefont{A.}~\bibnamefont{Pasupathy}},
  \bibinfo{author}{\bibfnamefont{J.}~\bibnamefont{Goldsmith}},
  \bibinfo{author}{\bibfnamefont{C.}~\bibnamefont{Chang}},
  \bibinfo{author}{\bibfnamefont{Y.}~\bibnamefont{Yaish}},
  \bibinfo{author}{\bibfnamefont{J.}~\bibnamefont{Petta}},
  \bibinfo{author}{\bibfnamefont{M.}~\bibnamefont{Rinkoski}},
  \bibinfo{author}{\bibfnamefont{J.}~\bibnamefont{Sethna}},
  \bibinfo{author}{\bibfnamefont{H.}~\bibnamefont{Abruna}},
  \bibinfo{author}{\bibfnamefont{P.}~\bibnamefont{McEuen}}, \bibnamefont{and}
  \bibinfo{author}{\bibfnamefont{D.}~\bibnamefont{Ralph}},
  \bibinfo{journal}{Nature (London)} \textbf{\bibinfo{volume}{417}},
  \bibinfo{pages}{722} (\bibinfo{year}{2002}).

\bibitem{Heersche06}
\bibinfo{author}{\bibfnamefont{H.~B.} \bibnamefont{Heersche}},
  \bibinfo{author}{\bibfnamefont{Z.}~\bibnamefont{de~Groot}},
  \bibinfo{author}{\bibfnamefont{J.~A.} \bibnamefont{Folk}},
  \bibinfo{author}{\bibfnamefont{H.~S.~J.} \bibnamefont{van~der Zant}},
  \bibinfo{author}{\bibfnamefont{C.}~\bibnamefont{Romeike}},
  \bibinfo{author}{\bibfnamefont{M.~R.} \bibnamefont{Wegewijs}},
  \bibinfo{author}{\bibfnamefont{L.}~\bibnamefont{Zobbi}},
  \bibinfo{author}{\bibfnamefont{D.}~\bibnamefont{Barreca}},
  \bibinfo{author}{\bibfnamefont{E.}~\bibnamefont{Tondello}}, \bibnamefont{and}
  \bibinfo{author}{\bibfnamefont{A.}~\bibnamefont{Cornia}},
  \bibinfo{journal}{Phys. Rev. Lett.} \textbf{\bibinfo{volume}{96}},
  \bibinfo{pages}{206801} (\bibinfo{year}{2006}).

\bibitem{lia02:725}
\bibinfo{author}{\bibfnamefont{W.}~\bibnamefont{Liang}},
  \bibinfo{author}{\bibfnamefont{M.}~\bibnamefont{Shores}},
  \bibinfo{author}{\bibfnamefont{M.}~\bibnamefont{Bockrath}},
  \bibinfo{author}{\bibfnamefont{J.}~\bibnamefont{Long}}, \bibnamefont{and}
  \bibinfo{author}{\bibfnamefont{H.}~\bibnamefont{Park}},
  \bibinfo{journal}{Nature (London)} \textbf{\bibinfo{volume}{417}},
  \bibinfo{pages}{725} (\bibinfo{year}{2002}).

\bibitem{Roch09}
\bibinfo{author}{\bibfnamefont{N.}~\bibnamefont{Roch}},
  \bibinfo{author}{\bibfnamefont{S.}~\bibnamefont{Florens}},
  \bibinfo{author}{\bibfnamefont{T.~A.} \bibnamefont{Costi}},
  \bibinfo{author}{\bibfnamefont{W.}~\bibnamefont{Wernsdorfer}},
  \bibnamefont{and} \bibinfo{author}{\bibfnamefont{F.}~\bibnamefont{Balestro}},
  \bibinfo{journal}{Phys. Rev. Lett.} \textbf{\bibinfo{volume}{103}},
  \bibinfo{pages}{197202} (\bibinfo{year}{2009}).

\bibitem{Parks10}
\bibinfo{author}{\bibfnamefont{J.~J.} \bibnamefont{Parks}},
  \bibinfo{author}{\bibfnamefont{A.~R.} \bibnamefont{Champagne}},
  \bibinfo{author}{\bibfnamefont{T.~A.} \bibnamefont{Costi}},
  \bibinfo{author}{\bibfnamefont{W.~W.} \bibnamefont{Shum}},
  \bibinfo{author}{\bibfnamefont{A.~N.} \bibnamefont{Pasupathy}},
  \bibinfo{author}{\bibfnamefont{E.}~\bibnamefont{Neuscamman}},
  \bibinfo{author}{\bibfnamefont{S.}~\bibnamefont{Flores-Torres}},
  \bibinfo{author}{\bibfnamefont{P.~S.} \bibnamefont{Cornaglia}},
  \bibinfo{author}{\bibfnamefont{A.~A.} \bibnamefont{Aligia}},
  \bibinfo{author}{\bibfnamefont{C.~A.} \bibnamefont{Balseiro}},
  \bibinfo{author}{\bibfnamefont{G.~K.-L.} \bibnamefont{Chan}},
  \bibinfo{author}{\bibfnamefont{H.~D.} \bibnamefont{Abru{\~n}a}},
  \emph{et~al.}, \bibinfo{journal}{Science} \textbf{\bibinfo{volume}{328}},
  \bibinfo{pages}{1370} (\bibinfo{year}{2010}).

\bibitem{Meded11}
\bibinfo{author}{\bibfnamefont{V.}~\bibnamefont{Meded}},
  \bibinfo{author}{\bibfnamefont{A.}~\bibnamefont{Bagrets}},
  \bibinfo{author}{\bibfnamefont{K.}~\bibnamefont{Fink}},
  \bibinfo{author}{\bibfnamefont{R.}~\bibnamefont{Chandrasekar}},
  \bibinfo{author}{\bibfnamefont{M.}~\bibnamefont{Ruben}},
  \bibinfo{author}{\bibfnamefont{F.}~\bibnamefont{Evers}},
  \bibinfo{author}{\bibfnamefont{A.}~\bibnamefont{Bernand-Mantel}},
  \bibinfo{author}{\bibfnamefont{J.~S.} \bibnamefont{Seldenthuis}},
  \bibinfo{author}{\bibfnamefont{A.}~\bibnamefont{Beukman}}, \bibnamefont{and}
  \bibinfo{author}{\bibfnamefont{H.~S.~J.} \bibnamefont{van~der Zant}},
  \bibinfo{journal}{Phys. Rev. B} \textbf{\bibinfo{volume}{83}},
  \bibinfo{pages}{245415} (\bibinfo{year}{2011}).

\bibitem{Esat16}
\bibinfo{author}{\bibfnamefont{T.}~\bibnamefont{Esat}},
  \bibinfo{author}{\bibfnamefont{B.}~\bibnamefont{Lechtenberg}},
  \bibinfo{author}{\bibfnamefont{T.}~\bibnamefont{Deilmann}},
  \bibinfo{author}{\bibfnamefont{C.}~\bibnamefont{Wagner}},
  \bibinfo{author}{\bibfnamefont{P.}~\bibnamefont{{Kr\"uger}}},
  \bibinfo{author}{\bibfnamefont{R.}~\bibnamefont{Temirov}},
  \bibinfo{author}{\bibfnamefont{M.}~\bibnamefont{Rohlfing}},
  \bibinfo{author}{\bibfnamefont{F.}~\bibnamefont{Anders}}, \bibnamefont{and}
  \bibinfo{author}{\bibfnamefont{F.}~\bibnamefont{Tautz}},
  \bibinfo{journal}{Nat. Phys.} \textbf{\bibinfo{volume}{12}},
  \bibinfo{pages}{867} (\bibinfo{year}{2016}).

\bibitem{che99:1550}
\bibinfo{author}{\bibfnamefont{J.}~\bibnamefont{Chen}},
  \bibinfo{author}{\bibfnamefont{M.}~\bibnamefont{Reed}},
  \bibinfo{author}{\bibfnamefont{A.}~\bibnamefont{Rawlett}}, \bibnamefont{and}
  \bibinfo{author}{\bibfnamefont{J.}~\bibnamefont{Tour}},
  \bibinfo{journal}{Science} \textbf{\bibinfo{volume}{286}},
  \bibinfo{pages}{1550} (\bibinfo{year}{1999}).

\bibitem{Gaudioso00}
\bibinfo{author}{\bibfnamefont{J.}~\bibnamefont{Gaudioso}},
  \bibinfo{author}{\bibfnamefont{L.~J.} \bibnamefont{Lauhon}},
  \bibnamefont{and} \bibinfo{author}{\bibfnamefont{W.}~\bibnamefont{Ho}},
  \bibinfo{journal}{Phys. Rev. Lett.} \textbf{\bibinfo{volume}{85}},
  \bibinfo{pages}{1918} (\bibinfo{year}{2000}).

\bibitem{Osorio10}
\bibinfo{author}{\bibfnamefont{E.~A.} \bibnamefont{Osorio}},
  \bibinfo{author}{\bibfnamefont{M.}~\bibnamefont{Ruben}},
  \bibinfo{author}{\bibfnamefont{J.~S.} \bibnamefont{Seldenthuis}},
  \bibinfo{author}{\bibfnamefont{J.~M.} \bibnamefont{Lehn}}, \bibnamefont{and}
  \bibinfo{author}{\bibfnamefont{H.~S.~J.} \bibnamefont{van~der Zant}},
  \bibinfo{journal}{Small} \textbf{\bibinfo{volume}{6}}, \bibinfo{pages}{174}
  (\bibinfo{year}{2010}).

\bibitem{blu05:167}
\bibinfo{author}{\bibfnamefont{A.}~\bibnamefont{Blum}},
  \bibinfo{author}{\bibfnamefont{J.}~\bibnamefont{Kushmerick}},
  \bibinfo{author}{\bibfnamefont{D.}~\bibnamefont{Long}},
  \bibinfo{author}{\bibfnamefont{C.}~\bibnamefont{Patterson}},
  \bibinfo{author}{\bibfnamefont{J.}~\bibnamefont{Jang}},
  \bibinfo{author}{\bibfnamefont{J.}~\bibnamefont{Henderson}},
  \bibinfo{author}{\bibfnamefont{Y.}~\bibnamefont{Yao}},
  \bibinfo{author}{\bibfnamefont{J.}~\bibnamefont{Tour}},
  \bibinfo{author}{\bibfnamefont{R.}~\bibnamefont{Shashidhar}},
  \bibnamefont{and} \bibinfo{author}{\bibfnamefont{B.}~\bibnamefont{Ratna}},
  \bibinfo{journal}{Nat. Mater.} \textbf{\bibinfo{volume}{4}},
  \bibinfo{pages}{167} (\bibinfo{year}{2005}).

\bibitem{Riel06}
\bibinfo{author}{\bibfnamefont{E.}~\bibnamefont{L\"ortscher}},
  \bibinfo{author}{\bibfnamefont{J.~W.} \bibnamefont{Ciszek}},
  \bibinfo{author}{\bibfnamefont{J.}~\bibnamefont{Tour}}, \bibnamefont{and}
  \bibinfo{author}{\bibfnamefont{H.}~\bibnamefont{Riel}},
  \bibinfo{journal}{Small} \textbf{\bibinfo{volume}{2}}, \bibinfo{pages}{973}
  (\bibinfo{year}{2006}).

\bibitem{Choi06}
\bibinfo{author}{\bibfnamefont{B.-Y.} \bibnamefont{Choi}},
  \bibinfo{author}{\bibfnamefont{S.-J.} \bibnamefont{Kahng}},
  \bibinfo{author}{\bibfnamefont{S.}~\bibnamefont{Kim}},
  \bibinfo{author}{\bibfnamefont{H.}~\bibnamefont{Kim}},
  \bibinfo{author}{\bibfnamefont{H.}~\bibnamefont{Kim}},
  \bibinfo{author}{\bibfnamefont{Y.}~\bibnamefont{Song}},
  \bibinfo{author}{\bibfnamefont{J.}~\bibnamefont{Ihm}}, \bibnamefont{and}
  \bibinfo{author}{\bibfnamefont{Y.}~\bibnamefont{Kuk}},
  \bibinfo{journal}{Phys. Rev. Lett.} \textbf{\bibinfo{volume}{96}},
  \bibinfo{pages}{156106} (\bibinfo{year}{2006}).

\bibitem{Bonca95}
\bibinfo{author}{\bibfnamefont{J.}~\bibnamefont{Bonca}} \bibnamefont{and}
  \bibinfo{author}{\bibfnamefont{S.}~\bibnamefont{Trugmann}},
  \bibinfo{journal}{Phys. Rev. Lett.} \textbf{\bibinfo{volume}{75}},
  \bibinfo{pages}{2566} (\bibinfo{year}{1995}).

\bibitem{Ness01}
\bibinfo{author}{\bibfnamefont{H.}~\bibnamefont{Ness}},
  \bibinfo{author}{\bibfnamefont{S.}~\bibnamefont{Shevlin}}, \bibnamefont{and}
  \bibinfo{author}{\bibfnamefont{A.}~\bibnamefont{Fisher}},
  \bibinfo{journal}{Phys. Rev. B} \textbf{\bibinfo{volume}{63}},
  \bibinfo{pages}{125422} (\bibinfo{year}{2001}).

\bibitem{Cizek04}
\bibinfo{author}{\bibfnamefont{M.}~\bibnamefont{Cizek}},
  \bibinfo{author}{\bibfnamefont{M.}~\bibnamefont{Thoss}}, \bibnamefont{and}
  \bibinfo{author}{\bibfnamefont{W.}~\bibnamefont{Domcke}},
  \bibinfo{journal}{Phys. Rev. B} \textbf{\bibinfo{volume}{70}},
  \bibinfo{pages}{125406} (\bibinfo{year}{2004}).

\bibitem{Cizek05}
\bibinfo{author}{\bibfnamefont{M.}~\bibnamefont{Cizek}},
  \bibinfo{author}{\bibfnamefont{M.}~\bibnamefont{Thoss}}, \bibnamefont{and}
  \bibinfo{author}{\bibfnamefont{W.}~\bibnamefont{Domcke}},
  \bibinfo{journal}{Czech.\ J.\ Phys.} \textbf{\bibinfo{volume}{55}},
  \bibinfo{pages}{189} (\bibinfo{year}{2005}).

\bibitem{Toroker07}
\bibinfo{author}{\bibfnamefont{M.}~\bibnamefont{Caspary-Toroker}}
  \bibnamefont{and} \bibinfo{author}{\bibfnamefont{U.}~\bibnamefont{Peskin}},
  \bibinfo{journal}{J. Chem. Phys.} \textbf{\bibinfo{volume}{127}},
  \bibinfo{pages}{154706} (\bibinfo{year}{2007}).

\bibitem{Benesch08}
\bibinfo{author}{\bibfnamefont{C.}~\bibnamefont{Benesch}},
  \bibinfo{author}{\bibfnamefont{M.}~\bibnamefont{Cizek}},
  \bibinfo{author}{\bibfnamefont{J.}~\bibnamefont{Klimes}},
  \bibinfo{author}{\bibfnamefont{I.}~\bibnamefont{Kondov}},
  \bibinfo{author}{\bibfnamefont{M.}~\bibnamefont{Thoss}}, \bibnamefont{and}
  \bibinfo{author}{\bibfnamefont{W.}~\bibnamefont{Domcke}},
  \bibinfo{journal}{J. Phys. Chem. C} \textbf{\bibinfo{volume}{112}},
  \bibinfo{pages}{9880} (\bibinfo{year}{2008}).

\bibitem{Zimbovskaya09}
\bibinfo{author}{\bibfnamefont{N.~A.} \bibnamefont{Zimbovskaya}}
  \bibnamefont{and} \bibinfo{author}{\bibfnamefont{M.~M.}
  \bibnamefont{Kuklja}}, \bibinfo{journal}{J. Chem. Phys.}
  \textbf{\bibinfo{volume}{131}}, \bibinfo{pages}{114703}
  (\bibinfo{year}{2009}).

\bibitem{Seidemann10}
\bibinfo{author}{\bibfnamefont{R.}~\bibnamefont{Jorn}} \bibnamefont{and}
  \bibinfo{author}{\bibfnamefont{T.}~\bibnamefont{Seidemann}},
  \bibinfo{journal}{J. Chem. Phys.} \textbf{\bibinfo{volume}{131}},
  \bibinfo{pages}{244114} (\bibinfo{year}{2009}).

\bibitem{Flensberg03}
\bibinfo{author}{\bibfnamefont{K.}~\bibnamefont{Flensberg}},
  \bibinfo{journal}{Phys. Rev. B} \textbf{\bibinfo{volume}{68}},
  \bibinfo{pages}{205323} (\bibinfo{year}{2003}).

\bibitem{Mitra04}
\bibinfo{author}{\bibfnamefont{A.}~\bibnamefont{Mitra}},
  \bibinfo{author}{\bibfnamefont{I.}~\bibnamefont{Aleiner}}, \bibnamefont{and}
  \bibinfo{author}{\bibfnamefont{A.~J.} \bibnamefont{Millis}},
  \bibinfo{journal}{Phys. Rev. B} \textbf{\bibinfo{volume}{69}},
  \bibinfo{pages}{245302} (\bibinfo{year}{2004}).

\bibitem{Galperin06}
\bibinfo{author}{\bibfnamefont{M.}~\bibnamefont{Galperin}},
  \bibinfo{author}{\bibfnamefont{M.}~\bibnamefont{Ratner}}, \bibnamefont{and}
  \bibinfo{author}{\bibfnamefont{A.}~\bibnamefont{Nitzan}},
  \bibinfo{journal}{Phys. Rev. B} \textbf{\bibinfo{volume}{73}},
  \bibinfo{pages}{045314} (\bibinfo{year}{2006}).

\bibitem{Ryndyk06}
\bibinfo{author}{\bibfnamefont{D.~A.} \bibnamefont{Ryndyk}},
  \bibinfo{author}{\bibfnamefont{M.}~\bibnamefont{Hartung}}, \bibnamefont{and}
  \bibinfo{author}{\bibfnamefont{G.}~\bibnamefont{Cuniberti}},
  \bibinfo{journal}{Phys. Rev. B} \textbf{\bibinfo{volume}{73}},
  \bibinfo{pages}{045420} (\bibinfo{year}{2006}).

\bibitem{Frederiksen07}
\bibinfo{author}{\bibfnamefont{T.}~\bibnamefont{Frederiksen}},
  \bibinfo{author}{\bibfnamefont{M.}~\bibnamefont{Paulsson}},
  \bibinfo{author}{\bibfnamefont{M.}~\bibnamefont{Brandbyge}},
  \bibnamefont{and} \bibinfo{author}{\bibfnamefont{A.}~\bibnamefont{Jauho}},
  \bibinfo{journal}{Phys. Rev. B} \textbf{\bibinfo{volume}{75}},
  \bibinfo{pages}{205413} (\bibinfo{year}{2007}).

\bibitem{Tahir08}
\bibinfo{author}{\bibfnamefont{M.}~\bibnamefont{Tahir}} \bibnamefont{and}
  \bibinfo{author}{\bibfnamefont{A.}~\bibnamefont{MacKinnon}},
  \bibinfo{journal}{Phys. Rev. B} \textbf{\bibinfo{volume}{77}},
  \bibinfo{pages}{224305} (\bibinfo{year}{2008}).

\bibitem{Haertle08}
\bibinfo{author}{\bibfnamefont{R.}~\bibnamefont{H{\"a}rtle}},
  \bibinfo{author}{\bibfnamefont{C.}~\bibnamefont{Benesch}}, \bibnamefont{and}
  \bibinfo{author}{\bibfnamefont{M.}~\bibnamefont{Thoss}},
  \bibinfo{journal}{Phys. Rev. B} \textbf{\bibinfo{volume}{77}},
  \bibinfo{pages}{205314} (\bibinfo{year}{2008}).

\bibitem{Bergfield09}
\bibinfo{author}{\bibfnamefont{J.~P.} \bibnamefont{Bergfield}}
  \bibnamefont{and} \bibinfo{author}{\bibfnamefont{C.~A.}
  \bibnamefont{Stafford}}, \bibinfo{journal}{Phys. Rev. B}
  \textbf{\bibinfo{volume}{79}}, \bibinfo{pages}{245125}
  (\bibinfo{year}{2009}).

\bibitem{Haertle09}
\bibinfo{author}{\bibfnamefont{R.}~\bibnamefont{H{\"a}rtle}},
  \bibinfo{author}{\bibfnamefont{C.}~\bibnamefont{Benesch}}, \bibnamefont{and}
  \bibinfo{author}{\bibfnamefont{M.}~\bibnamefont{Thoss}},
  \bibinfo{journal}{Phys. Rev. Lett.} \textbf{\bibinfo{volume}{102}},
  \bibinfo{pages}{146801} (\bibinfo{year}{2009}).

\bibitem{May02}
\bibinfo{author}{\bibfnamefont{V.}~\bibnamefont{May}}, \bibinfo{journal}{Phys.
  Rev. B} \textbf{\bibinfo{volume}{66}}, \bibinfo{pages}{245411}
  (\bibinfo{year}{2002}).

\bibitem{Lehmann04}
\bibinfo{author}{\bibfnamefont{J.}~\bibnamefont{Lehmann}},
  \bibinfo{author}{\bibfnamefont{S.}~\bibnamefont{Kohler}},
  \bibinfo{author}{\bibfnamefont{V.}~\bibnamefont{May}}, \bibnamefont{and}
  \bibinfo{author}{\bibfnamefont{P.}~\bibnamefont{{H\"anggi}}},
  \bibinfo{journal}{J. Chem. Phys.} \textbf{\bibinfo{volume}{121}},
  \bibinfo{pages}{2278} (\bibinfo{year}{2004}).

\bibitem{Pedersen05}
\bibinfo{author}{\bibfnamefont{J.~N.} \bibnamefont{Pedersen}} \bibnamefont{and}
  \bibinfo{author}{\bibfnamefont{A.}~\bibnamefont{Wacker}},
  \bibinfo{journal}{Phys. Rev. B} \textbf{\bibinfo{volume}{72}},
  \bibinfo{pages}{195330} (\bibinfo{year}{2005}).

\bibitem{Harbola06}
\bibinfo{author}{\bibfnamefont{U.}~\bibnamefont{Harbola}},
  \bibinfo{author}{\bibfnamefont{M.}~\bibnamefont{Esposito}}, \bibnamefont{and}
  \bibinfo{author}{\bibfnamefont{S.}~\bibnamefont{Mukamel}},
  \bibinfo{journal}{Phys. Rev. B} \textbf{\bibinfo{volume}{74}},
  \bibinfo{pages}{235309} (\bibinfo{year}{2006}).

\bibitem{Zazunov06}
\bibinfo{author}{\bibfnamefont{A.}~\bibnamefont{Zazunov}},
  \bibinfo{author}{\bibfnamefont{D.}~\bibnamefont{Feinberg}}, \bibnamefont{and}
  \bibinfo{author}{\bibfnamefont{T.}~\bibnamefont{Martin}},
  \bibinfo{journal}{Phys. Rev. B} \textbf{\bibinfo{volume}{73}},
  \bibinfo{pages}{115405} (\bibinfo{year}{2006}).

\bibitem{Siddiqui07}
\bibinfo{author}{\bibfnamefont{L.}~\bibnamefont{Siddiqui}},
  \bibinfo{author}{\bibfnamefont{A.~W.} \bibnamefont{Ghosh}}, \bibnamefont{and}
  \bibinfo{author}{\bibfnamefont{S.}~\bibnamefont{Datta}},
  \bibinfo{journal}{Phys. Rev. B} \textbf{\bibinfo{volume}{76}},
  \bibinfo{pages}{085433} (\bibinfo{year}{2007}).

\bibitem{Timm08}
\bibinfo{author}{\bibfnamefont{C.}~\bibnamefont{Timm}}, \bibinfo{journal}{Phys.
  Rev. B} \textbf{\bibinfo{volume}{77}}, \bibinfo{pages}{195416}
  (\bibinfo{year}{2008}).

\bibitem{May08}
\bibinfo{author}{\bibfnamefont{V.}~\bibnamefont{May}} \bibnamefont{and}
  \bibinfo{author}{\bibfnamefont{O.}~\bibnamefont{K\"uhn}},
  \bibinfo{journal}{Phys. Rev. B} \textbf{\bibinfo{volume}{77}},
  \bibinfo{pages}{115439} (\bibinfo{year}{2008}).

\bibitem{May08b}
\bibinfo{author}{\bibfnamefont{V.}~\bibnamefont{May}} \bibnamefont{and}
  \bibinfo{author}{\bibfnamefont{O.}~\bibnamefont{K\"uhn}},
  \bibinfo{journal}{Phys. Rev. B} \textbf{\bibinfo{volume}{77}},
  \bibinfo{pages}{115440} (\bibinfo{year}{2008}).

\bibitem{Leijnse08}
\bibinfo{author}{\bibfnamefont{M.}~\bibnamefont{Leijnse}} \bibnamefont{and}
  \bibinfo{author}{\bibfnamefont{M.~R.} \bibnamefont{Wegewijs}},
  \bibinfo{journal}{Phys. Rev. B} \textbf{\bibinfo{volume}{78}},
  \bibinfo{pages}{235424} (\bibinfo{year}{2008}).

\bibitem{Esposito09}
\bibinfo{author}{\bibfnamefont{M.}~\bibnamefont{Esposito}} \bibnamefont{and}
  \bibinfo{author}{\bibfnamefont{M.}~\bibnamefont{Galperin}},
  \bibinfo{journal}{Phys. Rev. B} \textbf{\bibinfo{volume}{79}},
  \bibinfo{pages}{205303} (\bibinfo{year}{2009}).

\bibitem{Volkovich11}
\bibinfo{author}{\bibfnamefont{R.}~\bibnamefont{Volkovich}},
  \bibinfo{author}{\bibfnamefont{R.}~\bibnamefont{{H\"artle}}},
  \bibinfo{author}{\bibfnamefont{M.}~\bibnamefont{Thoss}}, \bibnamefont{and}
  \bibinfo{author}{\bibfnamefont{U.}~\bibnamefont{Peskin}},
  \bibinfo{journal}{Phys. Chem. Chem. Phys.} \textbf{\bibinfo{volume}{13}},
  \bibinfo{pages}{14333} (\bibinfo{year}{2011}).

\bibitem{Haertle11}
\bibinfo{author}{\bibfnamefont{R.}~\bibnamefont{{H\"artle}}} \bibnamefont{and}
  \bibinfo{author}{\bibfnamefont{M.}~\bibnamefont{Thoss}},
  \bibinfo{journal}{Phys. Rev. B} \textbf{\bibinfo{volume}{83}},
  \bibinfo{pages}{115414} (\bibinfo{year}{2011}).

\bibitem{muh08:176403}
\bibinfo{author}{\bibfnamefont{L.}~\bibnamefont{{M\"uhlbacher}}}
  \bibnamefont{and} \bibinfo{author}{\bibfnamefont{E.}~\bibnamefont{Rabani}},
  \bibinfo{journal}{Phys. Rev. Lett.} \textbf{\bibinfo{volume}{100}},
  \bibinfo{pages}{176403} (\bibinfo{year}{2008}).

\bibitem{wei08:195316}
\bibinfo{author}{\bibfnamefont{S.}~\bibnamefont{Weiss}},
  \bibinfo{author}{\bibfnamefont{J.}~\bibnamefont{Eckel}},
  \bibinfo{author}{\bibfnamefont{M.}~\bibnamefont{Thorwart}}, \bibnamefont{and}
  \bibinfo{author}{\bibfnamefont{R.}~\bibnamefont{Egger}},
  \bibinfo{journal}{Phys. Rev. B} \textbf{\bibinfo{volume}{77}},
  \bibinfo{pages}{195316} (\bibinfo{year}{2008}).

\bibitem{Segal10}
\bibinfo{author}{\bibfnamefont{D.}~\bibnamefont{Segal}},
  \bibinfo{author}{\bibfnamefont{A.~J.} \bibnamefont{Millis}},
  \bibnamefont{and} \bibinfo{author}{\bibfnamefont{D.}~\bibnamefont{Reichman}},
  \bibinfo{journal}{Phys. Rev. B} \textbf{\bibinfo{volume}{82}},
  \bibinfo{pages}{205323} (\bibinfo{year}{2010}).

\bibitem{Werner09}
\bibinfo{author}{\bibfnamefont{P.}~\bibnamefont{Werner}},
  \bibinfo{author}{\bibfnamefont{T.}~\bibnamefont{Oka}}, \bibnamefont{and}
  \bibinfo{author}{\bibfnamefont{A.~J.} \bibnamefont{Millis}},
  \bibinfo{journal}{Phys. Rev. B} \textbf{\bibinfo{volume}{79}},
  \bibinfo{pages}{035320} (\bibinfo{year}{2009}).

\bibitem{Schiro09}
\bibinfo{author}{\bibfnamefont{M.}~\bibnamefont{Schiro}} \bibnamefont{and}
  \bibinfo{author}{\bibfnamefont{M.}~\bibnamefont{Fabrizio}},
  \bibinfo{journal}{Phys. Rev. B} \textbf{\bibinfo{volume}{79}},
  \bibinfo{pages}{153302} (\bibinfo{year}{2009}).

\bibitem{Cohen15}
\bibinfo{author}{\bibfnamefont{G.}~\bibnamefont{Cohen}},
  \bibinfo{author}{\bibfnamefont{E.}~\bibnamefont{Gull}},
  \bibinfo{author}{\bibfnamefont{D.}~\bibnamefont{Reichman}}, \bibnamefont{and}
  \bibinfo{author}{\bibfnamefont{A.}~\bibnamefont{Millis}},
  \bibinfo{journal}{Phys. Rev. Lett.} \textbf{\bibinfo{volume}{115}},
  \bibinfo{pages}{266802} (\bibinfo{year}{2015}).

\bibitem{and08:066804}
\bibinfo{author}{\bibfnamefont{F.~B.} \bibnamefont{Anders}},
  \bibinfo{journal}{Phys. Rev. Lett.} \textbf{\bibinfo{volume}{101}},
  \bibinfo{pages}{066804} (\bibinfo{year}{2008}).

\bibitem{HeidrichMeisner09}
\bibinfo{author}{\bibfnamefont{F.}~\bibnamefont{Heidrich-Meisner}},
  \bibinfo{author}{\bibfnamefont{A.}~\bibnamefont{Feiguin}}, \bibnamefont{and}
  \bibinfo{author}{\bibfnamefont{E.}~\bibnamefont{Dagotto}},
  \bibinfo{journal}{Phys. Rev. B} \textbf{\bibinfo{volume}{79}},
  \bibinfo{pages}{235336} (\bibinfo{year}{2009}).

\bibitem{Zheng09}
\bibinfo{author}{\bibfnamefont{X.}~\bibnamefont{Zheng}},
  \bibinfo{author}{\bibfnamefont{J.}~\bibnamefont{Jin}},
  \bibinfo{author}{\bibfnamefont{S.}~\bibnamefont{Welack}},
  \bibinfo{author}{\bibfnamefont{M.}~\bibnamefont{Luo}}, \bibnamefont{and}
  \bibinfo{author}{\bibfnamefont{Y.}~\bibnamefont{Yan}}, \bibinfo{journal}{J.
  Chem. Phys.} \textbf{\bibinfo{volume}{130}}, \bibinfo{pages}{164708}
  (\bibinfo{year}{2009}).

\bibitem{Jiang12}
\bibinfo{author}{\bibfnamefont{F.}~\bibnamefont{Jiang}},
  \bibinfo{author}{\bibfnamefont{J.}~\bibnamefont{Jin}},
  \bibinfo{author}{\bibfnamefont{S.}~\bibnamefont{Wang}}, \bibnamefont{and}
  \bibinfo{author}{\bibfnamefont{Y.}~\bibnamefont{Yan}},
  \bibinfo{journal}{Phys. Rev. B} \textbf{\bibinfo{volume}{85}},
  \bibinfo{pages}{245427} (\bibinfo{year}{2012}).

\bibitem{Zheng13}
\bibinfo{author}{\bibfnamefont{X.}~\bibnamefont{Zheng}},
  \bibinfo{author}{\bibfnamefont{Y.}~\bibnamefont{Yan}}, \bibnamefont{and}
  \bibinfo{author}{\bibfnamefont{M.~D.} \bibnamefont{Ventra}},
  \bibinfo{journal}{Phys. Rev. Lett.} \textbf{\bibinfo{volume}{111}},
  \bibinfo{pages}{086601} (\bibinfo{year}{2013}).

\bibitem{Haertle13c}
\bibinfo{author}{\bibfnamefont{R.}~\bibnamefont{{H\"artle}}},
  \bibinfo{author}{\bibfnamefont{G.}~\bibnamefont{Cohen}},
  \bibinfo{author}{\bibfnamefont{D.}~\bibnamefont{Reichman}}, \bibnamefont{and}
  \bibinfo{author}{\bibfnamefont{A.}~\bibnamefont{Millis}},
  \bibinfo{journal}{Phys. Rev. B} \textbf{\bibinfo{volume}{88}},
  \bibinfo{pages}{235426} (\bibinfo{year}{2013}).

\bibitem{Haertle14}
\bibinfo{author}{\bibfnamefont{R.}~\bibnamefont{{H\"artle}}} \bibnamefont{and}
  \bibinfo{author}{\bibfnamefont{A.}~\bibnamefont{Millis}},
  \bibinfo{journal}{Phys. Rev. B} \textbf{\bibinfo{volume}{90}},
  \bibinfo{pages}{075409} (\bibinfo{year}{2014}).

\bibitem{Schinabeck16}
\bibinfo{author}{\bibfnamefont{C.}~\bibnamefont{Schinabeck}},
  \bibinfo{author}{\bibfnamefont{R.}~\bibnamefont{{H\"artle}}},
  \bibnamefont{and} \bibinfo{author}{\bibfnamefont{M.}~\bibnamefont{Thoss}},
  \bibinfo{journal}{Phys. Rev. B} \textbf{\bibinfo{volume}{94}},
  \bibinfo{pages}{201407(R)} (\bibinfo{year}{2016}).

\bibitem{Cohen11}
\bibinfo{author}{\bibfnamefont{G.}~\bibnamefont{Cohen}} \bibnamefont{and}
  \bibinfo{author}{\bibfnamefont{E.}~\bibnamefont{Rabani}},
  \bibinfo{journal}{Phys. Rev. B} \textbf{\bibinfo{volume}{84}},
  \bibinfo{pages}{075150} (\bibinfo{year}{2011}).

\bibitem{Wilner15}
\bibinfo{author}{\bibfnamefont{E.}~\bibnamefont{Wilner}},
  \bibinfo{author}{\bibfnamefont{H.}~\bibnamefont{Wang}},
  \bibinfo{author}{\bibfnamefont{M.}~\bibnamefont{Thoss}}, \bibnamefont{and}
  \bibinfo{author}{\bibfnamefont{E.}~\bibnamefont{Rabani}},
  \bibinfo{journal}{Phys. Rev. B} \textbf{\bibinfo{volume}{92}},
  \bibinfo{pages}{195143} (\bibinfo{year}{2015}).

\bibitem{wan09:024114}
\bibinfo{author}{\bibfnamefont{H.}~\bibnamefont{Wang}} \bibnamefont{and}
  \bibinfo{author}{\bibfnamefont{M.}~\bibnamefont{Thoss}}, \bibinfo{journal}{J.
  Chem. Phys.} \textbf{\bibinfo{volume}{131}}, \bibinfo{pages}{024114}
  (\bibinfo{year}{2009}).

\bibitem{Wang11}
\bibinfo{author}{\bibfnamefont{H.}~\bibnamefont{Wang}},
  \bibinfo{author}{\bibfnamefont{I.}~\bibnamefont{Pshenichnyuk}},
  \bibinfo{author}{\bibfnamefont{R.}~\bibnamefont{{H\"artle}}},
  \bibnamefont{and} \bibinfo{author}{\bibfnamefont{M.}~\bibnamefont{Thoss}},
  \bibinfo{journal}{J. Chem. Phys.} \textbf{\bibinfo{volume}{135}},
  \bibinfo{pages}{244506} (\bibinfo{year}{2011}).

\bibitem{Wang13}
\bibinfo{author}{\bibfnamefont{H.}~\bibnamefont{Wang}} \bibnamefont{and}
  \bibinfo{author}{\bibfnamefont{M.}~\bibnamefont{Thoss}}, \bibinfo{journal}{J.
  Chem. Phys.} \textbf{\bibinfo{volume}{138}}, \bibinfo{pages}{134704}
  (\bibinfo{year}{2013}).

\bibitem{Wang16}
\bibinfo{author}{\bibfnamefont{H.}~\bibnamefont{Wang}} \bibnamefont{and}
  \bibinfo{author}{\bibfnamefont{M.}~\bibnamefont{Thoss}}, \bibinfo{journal}{J.
  Chem. Phys.} \textbf{\bibinfo{volume}{145}}, \bibinfo{pages}{164105}
  (\bibinfo{year}{2016}).

\bibitem{Wang13b}
\bibinfo{author}{\bibfnamefont{H.}~\bibnamefont{Wang}} \bibnamefont{and}
  \bibinfo{author}{\bibfnamefont{M.}~\bibnamefont{Thoss}}, \bibinfo{journal}{J.
  Phys. Chem. A} \textbf{\bibinfo{volume}{117}}, \bibinfo{pages}{7431}
  (\bibinfo{year}{2013}).

\bibitem{Hewson93}
\bibinfo{author}{\bibfnamefont{A.}~\bibnamefont{Hewson}},
  \emph{\bibinfo{title}{The Kondo Problem to Heavy Fermions}}
  (\bibinfo{publisher}{Cambridge Press}, \bibinfo{address}{Cambridge UK},
  \bibinfo{year}{1993}).

\bibitem{Wiel00}
\bibinfo{author}{\bibfnamefont{W.}~\bibnamefont{van~der Wiel}},
  \bibinfo{author}{\bibfnamefont{S.}~\bibnamefont{Franceschi}},
  \bibinfo{author}{\bibfnamefont{T.}~\bibnamefont{Fujisawa}},
  \bibinfo{author}{\bibfnamefont{J.}~\bibnamefont{Elzerman}},
  \bibinfo{author}{\bibfnamefont{S.}~\bibnamefont{Tarucha}}, \bibnamefont{and}
  \bibinfo{author}{\bibfnamefont{L.}~\bibnamefont{Kouwenhoven}},
  \bibinfo{journal}{Science} \textbf{\bibinfo{volume}{289}},
  \bibinfo{pages}{2105} (\bibinfo{year}{2000}).

\bibitem{Wilson75}
\bibinfo{author}{\bibfnamefont{K.}~\bibnamefont{Wilson}},
  \bibinfo{journal}{Rev. Mod. Phys.} \textbf{\bibinfo{volume}{47}},
  \bibinfo{pages}{773} (\bibinfo{year}{1975}).

\bibitem{Anderson61}
\bibinfo{author}{\bibfnamefont{P.}~\bibnamefont{Anderson}},
  \bibinfo{journal}{Phys. Rev.} \textbf{\bibinfo{volume}{124}},
  \bibinfo{pages}{41} (\bibinfo{year}{1961}).

\bibitem{Eckel10}
\bibinfo{author}{\bibfnamefont{J.}~\bibnamefont{Eckel}},
  \bibinfo{author}{\bibfnamefont{F.}~\bibnamefont{Heidrich-Meisner}},
  \bibinfo{author}{\bibfnamefont{S.}~\bibnamefont{Jakobs}},
  \bibinfo{author}{\bibfnamefont{M.}~\bibnamefont{Thorwart}},
  \bibinfo{author}{\bibfnamefont{M.}~\bibnamefont{Pletyukhov}},
  \bibnamefont{and} \bibinfo{author}{\bibfnamefont{R.}~\bibnamefont{Egger}},
  \bibinfo{journal}{New. J. Phys.} \textbf{\bibinfo{volume}{12}},
  \bibinfo{pages}{043042} (\bibinfo{year}{2010}).

\bibitem{Benesch09}
\bibinfo{author}{\bibfnamefont{C.}~\bibnamefont{Benesch}},
  \bibinfo{author}{\bibfnamefont{M.}~\bibnamefont{Rode}},
  \bibinfo{author}{\bibfnamefont{M.}~\bibnamefont{Cizek}},
  \bibinfo{author}{\bibfnamefont{R.}~\bibnamefont{{H\"artle}}},
  \bibinfo{author}{\bibfnamefont{O.}~\bibnamefont{Rubio-Pons}},
  \bibinfo{author}{\bibfnamefont{M.}~\bibnamefont{Thoss}}, \bibnamefont{and}
  \bibinfo{author}{\bibfnamefont{A.}~\bibnamefont{Sobolewski}},
  \bibinfo{journal}{J. Phys. Chem. C} \textbf{\bibinfo{volume}{112}},
  \bibinfo{pages}{9880} (\bibinfo{year}{2008}).

\bibitem{wan06:034114}
\bibinfo{author}{\bibfnamefont{H.}~\bibnamefont{Wang}} \bibnamefont{and}
  \bibinfo{author}{\bibfnamefont{M.}~\bibnamefont{Thoss}}, \bibinfo{journal}{J.
  Chem. Phys.} \textbf{\bibinfo{volume}{124}}(\bibinfo{number}{3}),
  \bibinfo{pages}{034114} (\bibinfo{year}{2006}).

\bibitem{tho01:2991}
\bibinfo{author}{\bibfnamefont{M.}~\bibnamefont{Thoss}},
  \bibinfo{author}{\bibfnamefont{H.}~\bibnamefont{Wang}}, \bibnamefont{and}
  \bibinfo{author}{\bibfnamefont{W.~H.} \bibnamefont{Miller}},
  \bibinfo{journal}{J. Chem. Phys.}
  \textbf{\bibinfo{volume}{115}}(\bibinfo{number}{7}), \bibinfo{pages}{2991}
  (\bibinfo{year}{2001}).

\bibitem{wan01:2979}
\bibinfo{author}{\bibfnamefont{H.}~\bibnamefont{Wang}},
  \bibinfo{author}{\bibfnamefont{M.}~\bibnamefont{Thoss}}, \bibnamefont{and}
  \bibinfo{author}{\bibfnamefont{W.~H.} \bibnamefont{Miller}},
  \bibinfo{journal}{J. Chem. Phys.}
  \textbf{\bibinfo{volume}{115}}(\bibinfo{number}{7}), \bibinfo{pages}{2979}
  (\bibinfo{year}{2001}).

\bibitem{wan03:1289}
\bibinfo{author}{\bibfnamefont{H.}~\bibnamefont{Wang}} \bibnamefont{and}
  \bibinfo{author}{\bibfnamefont{M.}~\bibnamefont{Thoss}}, \bibinfo{journal}{J.
  Chem. Phys.} \textbf{\bibinfo{volume}{119}}(\bibinfo{number}{3}),
  \bibinfo{pages}{1289} (\bibinfo{year}{2003}).

\bibitem{Goldberg1978}
\bibinfo{author}{\bibfnamefont{A.}~\bibnamefont{Goldberg}} \bibnamefont{and}
  \bibinfo{author}{\bibfnamefont{B.~W.} \bibnamefont{Shore}},
  \bibinfo{journal}{J. Phys. B} \textbf{\bibinfo{volume}{11}},
  \bibinfo{pages}{3339} (\bibinfo{year}{1978}).

\bibitem{Kosloff1986}
\bibinfo{author}{\bibfnamefont{R.}~\bibnamefont{Kosloff}} \bibnamefont{and}
  \bibinfo{author}{\bibfnamefont{D.}~\bibnamefont{Kosloff}},
  \bibinfo{journal}{J. Comput. Phys.} \textbf{\bibinfo{volume}{63}},
  \bibinfo{pages}{363} (\bibinfo{year}{1986}).

\bibitem{Neuhauser1989}
\bibinfo{author}{\bibfnamefont{D.}~\bibnamefont{Neuhauser}} \bibnamefont{and}
  \bibinfo{author}{\bibfnamefont{M.}~\bibnamefont{Baer}}, \bibinfo{journal}{J.
  Chem. Phys.} \textbf{\bibinfo{volume}{90}}, \bibinfo{pages}{4351}
  (\bibinfo{year}{1989}).

\bibitem{Seideman1991}
\bibinfo{author}{\bibfnamefont{T.}~\bibnamefont{Seideman}} \bibnamefont{and}
  \bibinfo{author}{\bibfnamefont{W.~H.} \bibnamefont{Miller}},
  \bibinfo{journal}{J. Chem. Phys.} \textbf{\bibinfo{volume}{96}},
  \bibinfo{pages}{4412} (\bibinfo{year}{1991}).

\bibitem{wan10:78}
\bibinfo{author}{\bibfnamefont{H.}~\bibnamefont{Wang}} \bibnamefont{and}
  \bibinfo{author}{\bibfnamefont{M.}~\bibnamefont{Thoss}},
  \bibinfo{journal}{Chem. Phys.} \textbf{\bibinfo{volume}{370}},
  \bibinfo{pages}{78} (\bibinfo{year}{2010}).

\bibitem{Wang15}
\bibinfo{author}{\bibfnamefont{H.}~\bibnamefont{Wang}}, \bibinfo{journal}{J.
  Phys. Chem. A} \textbf{\bibinfo{volume}{119}}, \bibinfo{pages}{7951}
  (\bibinfo{year}{2015}).

\bibitem{Frenkel34}
\bibinfo{author}{\bibfnamefont{J.}~\bibnamefont{Frenkel}},
  \emph{\bibinfo{title}{Wave Mechanics}} (\bibinfo{publisher}{Clarendon Press},
  \bibinfo{address}{Oxford}, \bibinfo{year}{1934}).

\bibitem{mey90:73}
\bibinfo{author}{\bibfnamefont{H.-D.} \bibnamefont{Meyer}},
  \bibinfo{author}{\bibfnamefont{U.}~\bibnamefont{Manthe}}, \bibnamefont{and}
  \bibinfo{author}{\bibfnamefont{L.~S.} \bibnamefont{Cederbaum}},
  \bibinfo{journal}{Chem. Phys. Lett.}
  \textbf{\bibinfo{volume}{165}}(\bibinfo{number}{1}), \bibinfo{pages}{73}
  (\bibinfo{year}{1990}).

\bibitem{man92:3199}
\bibinfo{author}{\bibfnamefont{U.}~\bibnamefont{Manthe}},
  \bibinfo{author}{\bibfnamefont{H.-D.} \bibnamefont{Meyer}}, \bibnamefont{and}
  \bibinfo{author}{\bibfnamefont{L.~S.} \bibnamefont{Cederbaum}},
  \bibinfo{journal}{J.~Chem.\ Phys.} \textbf{\bibinfo{volume}{97}},
  \bibinfo{pages}{3199} (\bibinfo{year}{1992}).

\bibitem{bec00:1}
\bibinfo{author}{\bibfnamefont{M.~H.} \bibnamefont{Beck}},
  \bibinfo{author}{\bibfnamefont{A.}~\bibnamefont{J\"ackle}},
  \bibinfo{author}{\bibfnamefont{G.~A.} \bibnamefont{Worth}}, \bibnamefont{and}
  \bibinfo{author}{\bibfnamefont{H.~.-D.} \bibnamefont{Meyer}},
  \bibinfo{journal}{Phys. Rep.}
  \textbf{\bibinfo{volume}{324}}(\bibinfo{number}{1}), \bibinfo{pages}{1}
  (\bibinfo{year}{2000}).

\bibitem{mey03:251}
\bibinfo{author}{\bibfnamefont{H.-D.} \bibnamefont{Meyer}} \bibnamefont{and}
  \bibinfo{author}{\bibfnamefont{G.~A.} \bibnamefont{Worth}},
  \bibinfo{journal}{Theor.\ Chem.\ Acc.} \textbf{\bibinfo{volume}{109}},
  \bibinfo{pages}{251} (\bibinfo{year}{2003}).

\bibitem{Meyer09}
\bibinfo{author}{\bibfnamefont{H.-D.} \bibnamefont{Meyer}},
  \bibinfo{author}{\bibfnamefont{F.}~\bibnamefont{Gatti}}, \bibnamefont{and}
  \bibinfo{author}{\bibfnamefont{G.}~\bibnamefont{Worth}},
  \emph{\bibinfo{title}{Multidimensional Quantum Dynamics: MCTDH Theory and
  Applications}} (\bibinfo{publisher}{Whiley-VCH}, \bibinfo{address}{Weinheim},
  \bibinfo{year}{2009}).

\bibitem{Thoss04b}
\bibinfo{author}{\bibfnamefont{M.}~\bibnamefont{Thoss}},
  \bibinfo{author}{\bibfnamefont{I.}~\bibnamefont{Kondov}}, \bibnamefont{and}
  \bibinfo{author}{\bibfnamefont{H.}~\bibnamefont{Wang}},
  \bibinfo{journal}{Chem. Phys.} \textbf{\bibinfo{volume}{304}},
  \bibinfo{pages}{169} (\bibinfo{year}{2004}).

\bibitem{tho06:210}
\bibinfo{author}{\bibfnamefont{M.}~\bibnamefont{Thoss}} \bibnamefont{and}
  \bibinfo{author}{\bibfnamefont{H.}~\bibnamefont{Wang}},
  \bibinfo{journal}{Chem. Phys.}
  \textbf{\bibinfo{volume}{322}}(\bibinfo{number}{1-2}), \bibinfo{pages}{210}
  (\bibinfo{year}{2006}).

\bibitem{wan07:10369}
\bibinfo{author}{\bibfnamefont{H.}~\bibnamefont{Wang}} \bibnamefont{and}
  \bibinfo{author}{\bibfnamefont{M.}~\bibnamefont{Thoss}}, \bibinfo{journal}{J.
  Phys. Chem. A} \textbf{\bibinfo{volume}{111}}, \bibinfo{pages}{10369}
  (\bibinfo{year}{2007}).

\bibitem{cra07:144503}
\bibinfo{author}{\bibfnamefont{I.~R.} \bibnamefont{Craig}},
  \bibinfo{author}{\bibfnamefont{M.}~\bibnamefont{Thoss}}, \bibnamefont{and}
  \bibinfo{author}{\bibfnamefont{H.}~\bibnamefont{Wang}}, \bibinfo{journal}{J.
  Chem. Phys.} \textbf{\bibinfo{volume}{127}}, \bibinfo{pages}{144503}
  (\bibinfo{year}{2007}).

\bibitem{vel08:325}
\bibinfo{author}{\bibfnamefont{K.~A.} \bibnamefont{Velizhanin}},
  \bibinfo{author}{\bibfnamefont{H.}~\bibnamefont{Wang}}, \bibnamefont{and}
  \bibinfo{author}{\bibfnamefont{M.}~\bibnamefont{Thoss}},
  \bibinfo{journal}{Chem. Phys. Lett.} \textbf{\bibinfo{volume}{460}},
  \bibinfo{pages}{325} (\bibinfo{year}{2008}).

\bibitem{wan06:174502}
\bibinfo{author}{\bibfnamefont{H.}~\bibnamefont{Wang}},
  \bibinfo{author}{\bibfnamefont{D.}~\bibnamefont{Skinner}}, \bibnamefont{and}
  \bibinfo{author}{\bibfnamefont{M.}~\bibnamefont{Thoss}}, \bibinfo{journal}{J.
  Chem. Phys.} \textbf{\bibinfo{volume}{125}}, \bibinfo{pages}{174502}
  (\bibinfo{year}{2006}).

\bibitem{man08:164116}
\bibinfo{author}{\bibfnamefont{U.}~\bibnamefont{Manthe}}, \bibinfo{journal}{J.
  Chem. Phys.} \textbf{\bibinfo{volume}{128}}, \bibinfo{pages}{164116}
  (\bibinfo{year}{2008}).

\bibitem{man09:054109}
\bibinfo{author}{\bibfnamefont{U.}~\bibnamefont{Manthe}}, \bibinfo{journal}{J.
  Chem. Phys.} \textbf{\bibinfo{volume}{130}}, \bibinfo{pages}{054109}
  (\bibinfo{year}{2009}).

\bibitem{Wang17}
\bibinfo{author}{\bibfnamefont{H.}~\bibnamefont{Wang}} \bibnamefont{and}
  \bibinfo{author}{\bibfnamefont{M.}~\bibnamefont{Thoss}}, \bibinfo{journal}{J.
  Chem. Phys.} \textbf{\bibinfo{volume}{146}}, \bibinfo{pages}{124112}
  (\bibinfo{year}{2017}).

\bibitem{alo08:033613}
\bibinfo{author}{\bibfnamefont{O.~E.} \bibnamefont{Alon}},
  \bibinfo{author}{\bibfnamefont{A.~I.} \bibnamefont{Streltsov}},
  \bibnamefont{and} \bibinfo{author}{\bibfnamefont{L.~S.}
  \bibnamefont{Cederbaum}}, \bibinfo{journal}{Phys. Rev. A}
  \textbf{\bibinfo{volume}{77}}, \bibinfo{pages}{033613}
  (\bibinfo{year}{2008}).

\bibitem{kat04:533}
\bibinfo{author}{\bibfnamefont{T.}~\bibnamefont{Kato}} \bibnamefont{and}
  \bibinfo{author}{\bibfnamefont{H.}~\bibnamefont{Kono}},
  \bibinfo{journal}{Chem. Phys. Lett.} \textbf{\bibinfo{volume}{392}},
  \bibinfo{pages}{533} (\bibinfo{year}{2004}).

\bibitem{cai05:012712}
\bibinfo{author}{\bibfnamefont{J.}~\bibnamefont{Caillat}},
  \bibinfo{author}{\bibfnamefont{J.}~\bibnamefont{Zanghellini}},
  \bibinfo{author}{\bibfnamefont{M.}~\bibnamefont{Kitzler}},
  \bibinfo{author}{\bibfnamefont{O.}~\bibnamefont{Koch}},
  \bibinfo{author}{\bibfnamefont{W.}~\bibnamefont{Kreuzer}}, \bibnamefont{and}
  \bibinfo{author}{\bibfnamefont{A.}~\bibnamefont{Scrinzi}},
  \bibinfo{journal}{Phys. Rev. A} \textbf{\bibinfo{volume}{71}},
  \bibinfo{pages}{012712} (\bibinfo{year}{2005}).

\bibitem{nes05:124102}
\bibinfo{author}{\bibfnamefont{M.}~\bibnamefont{Nest}},
  \bibinfo{author}{\bibfnamefont{T.}~\bibnamefont{Klamroth}}, \bibnamefont{and}
  \bibinfo{author}{\bibfnamefont{P.}~\bibnamefont{Saalfrank}},
  \bibinfo{journal}{J. Chem. Phys.} \textbf{\bibinfo{volume}{122}},
  \bibinfo{pages}{124102} (\bibinfo{year}{2005}).

\bibitem{Manthe17}
\bibinfo{author}{\bibfnamefont{U.}~\bibnamefont{Manthe}} \bibnamefont{and}
  \bibinfo{author}{\bibfnamefont{T.}~\bibnamefont{Weike}}, \bibinfo{journal}{J.
  Chem. Phys.} \textbf{\bibinfo{volume}{146}}, \bibinfo{pages}{064117}
  (\bibinfo{year}{2017}).

\bibitem{Balzer15}
\bibinfo{author}{\bibfnamefont{K.}~\bibnamefont{Balzer}},
  \bibinfo{author}{\bibfnamefont{Z.}~\bibnamefont{Li}},
  \bibinfo{author}{\bibfnamefont{O.}~\bibnamefont{Vendrell}}, \bibnamefont{and}
  \bibinfo{author}{\bibfnamefont{M.}~\bibnamefont{Eckstein}},
  \bibinfo{journal}{Phys. Rev. B} \textbf{\bibinfo{volume}{91}},
  \bibinfo{pages}{2015} (\bibinfo{year}{045136}).

\bibitem{Oguri05}
\bibinfo{author}{\bibfnamefont{A.}~\bibnamefont{Oguri}}, \bibinfo{journal}{J.
  Phys. Soc. Jpn.} \textbf{\bibinfo{volume}{74}}, \bibinfo{pages}{110}
  (\bibinfo{year}{2005}).

\end{thebibliography}
\end{document}